\documentclass[12pt,preprint]{aastex}

\usepackage{fullpage}

\newcommand{\doubint}{\int\!\!\!\!\int}
\newcommand{\bK}{ \mbox{\bf K} }
\newcommand{\ii}{ {\rm i} }
\newcommand{\id}{ {\rm d} }

\newcommand{\bk}{ \mbox{\boldmath$k$} }

\newcommand{\cC}{ \mbox{\boldmath$\cal C$} }

\newcommand{\cJ}{ {\cal J} }
\newcommand{\cJJ}{ \mbox{\boldmath$\cal J$} }

\newcommand{\Cref}{ {C_{\rm ref}} }
\newcommand{\bu}{ \mbox{\boldmath$u$} }

\newcommand{\by}{ \mbox{\boldmath$y$} }
\newcommand{\bDelta}{ \mbox{\boldmath$\Delta$} }
\newcommand{\los}{ \mbox{\boldmath$\ell$} }
\newcommand{\1}{{\mbox{\boldmath$x_1$}}}
\newcommand{\2}{{\mbox{\boldmath$x_2$}}}

\newcommand{\bx}{{\mbox{\boldmath$x$}}}
\newcommand{\br}{{\mbox{\boldmath$r$}}}
\newcommand{\bz}{{\mbox{\boldmath$z$}}}

\newcommand{\zhat}{ \mbox{\boldmath$\hat{z}$} }
\newcommand{\bbv}{ \mbox{\boldmath$v$} }

\newcommand{\bnabla}{ \mbox{\boldmath$\nabla$} }

\newcommand{\sbx}{{\mbox{\scriptsize\boldmath$x$}}}
\newcommand{\sbk}{{\mbox{\scriptsize\boldmath$k$}}}

\newcommand{\sbDelta}{{\mbox{\scriptsize\boldmath$\Delta$}}}

\begin{document}
\shorttitle{Time-distance helioseismology}
\shortauthors{Jackiewicz et al.}
\title{Time-distance helioseismology: Sensitivity of f-mode travel times to flows}
\author{J. Jackiewicz}
\affil{Max-Planck-Institut f\"{u}r Sonnensystemforschung, 37191 Katlenburg-Lindau, Germany}
\email{jackiewicz@mps.mpg.de}
\author{L. Gizon}
\affil{Max-Planck-Institut f\"{u}r Sonnensystemforschung, 37191 Katlenburg-Lindau, Germany}
\author{A.~C. Birch}
\affil{NWRA, CoRA Division, 3380 Mitchell Lane, Boulder, CO 80301}
\author{T.~L. Duvall Jr.}
\affil{Laboratory for Astronomy and Solar Physics, NASA Goddard Space Flight Center, Greenbelt, MD 20771}

\begin{abstract}
Time-distance helioseismology has shown that f-mode travel times contain information about horizontal flows in the Sun. The purpose of this study is to provide a simple interpretation of these travel times. We study the interaction of surface-gravity waves with horizontal flows in an incompressible, plane-parallel solar atmosphere. We show that for uniform flows less than roughly 250~m\,s$^{-1}$, the travel-time shifts are linear in the flow amplitude. For stronger flows, perturbation theory up to third order is needed to model waveforms. The case of small-amplitude  spatially-varying flows is treated using the first-order Born approximation. We derive two-dimensional Fr\'{e}chet kernels that give the sensitivity of travel-time shifts to local flows. We show that the effect of flows on travel times depends on wave damping and on the direction from which the observations are made.  The main physical effect is the advection of the waves by the flow rather than the advection of wave sources or the effect of flows on wave damping.  We compare the two-dimensional sensitivity kernels with simplified three-dimensional kernels that only account for wave advection and assume a vertical line of sight. We find that the three-dimensional f-mode kernels approximately separate in the horizontal and vertical coordinates, with the horizontal variations given by the simplified two-dimensional kernels. This consistency between quite different models gives us confidence in the usefulness of these kernels for interpreting quiet-Sun observations. 
\end{abstract}


\section{Introduction}
One of the tasks of time-distance helioseismology \citep{duvall1993} is to map the flow field in the solar convection zone at the highest spatial and temporal resolutions possible.  Using a cross-correlation technique, it is possible to estimate the time taken by waves to travel from one point on the solar surface to another.  These travel times contain information about local flows \citep[e.g.][]{duvall2000,gizon2000}. Particularly interesting are flows associated with supergranulation, active regions and sunspots, and connected with the evolving magnetic cycle of the Sun. The devil is in the details of how the flow information is encoded in and extracted from the travel-time measurements.

We study the forward problem in the case of solar surface-gravity waves, also called f modes, which propagate horizontally. For frequencies near 3~mHz the f-mode kinetic energy peaks about 1~Mm below the solar surface.  The interaction of f modes with flows has been studied by \citet{murawski2000}, \citet{duvall2000}, \citet{woodard2002}, \citet{zhao2007}, \citet{birch2007}, and \citet{birch2007ring}. Small-scale turbulent convection (granulation) excites and damps solar oscillations \citep[e.g.,][]{murawski2000}. \citet{birch2007} and \citet{birch2007ring} used the Born approximation \citep[e.g.][]{gizon2002} to compute the effects of the advection of waves by small-amplitude steady flows on time-distance and ring-diagram measurements.

Here we are interested in studying in more detail the sensitivity of f-mode travel times, as measured by time-distance helioseismology, to horizontal steady flows.  We wish to answer the following questions, none of which have been studied before. How do travel-time shifts depend on travel distance? In particular, what is the role of wave damping? Do travel-time shifts depend linearly on the flow amplitude for typical solar flows? If so, when does the linearity break down? Which physical effects contribute to the travel-time perturbations? Is advection of the waves by the flow the only important effect? Or is the Doppler shift of wave sources (granulation) by larger-scale flows also relevant? Is it important to model damping in a frame moving with the flow? In the limit of weak flows, we can obtain kernel functions that give the linear sensitivity of travel-time perturbations to spatially-varying horizontal flows. How do sensitivity kernels change with the bandwidth and the central frequency of the f-mode wave packets?  Waves at the solar surface are typically observed in line-of-sight velocity using Doppler measurements.  What effect does the projection of the wave velocity onto the line of sight have on the kernels? Can we find an accurate approximation to the kernels to speed up demanding calculations?  Is it possible to recover the ray approximation of \citet{kosovichev1997} from the high-frequency limit of the cross-path integral of the sensitivity kernels \citep[see][]{tong1998}? 

In order to answer these questions, we choose to work with a simplified model of surface-gravity waves. Following \citet{gizon2002}, we consider waves at the free surface of a constant-density half space under constant gravity. We choose initial conditions such that the wave field remains irrotational.  These assumptions are useful as they simplify the presentation; they are also sensible as will be shown at the end of the paper through a comparison with a solar-like three-dimensional kernel calculation with otherwise simplified physics.

The layout of the rest of the paper is as follows. In \S~\ref{sec:surfacewaves} we introduce our simple surface-gravity wave model. In \S~\ref{sec:uniform} we study the effect of a uniform flow on wave travel times, finding an exact solution and one  using perturbation theory.  In \S~\ref{sec:generalflow} we use the first-order Born approximation to obtain the wave field in the case of a small-amplitude spatially-varying flow, and then derive the sensitivity kernels. \S~\ref{sec:3D} is a comparison with a three-dimensional kernel calculation. The implications of our findings are discussed in \S~\ref{sec:discussion}.


\section{A simplified model for f modes}
\label{sec:surfacewaves}

Following \citet{gizon2002}, we consider a constant-density, incompressible half space $z<0$, bounded by a free surface at $z=0$.  Throughout the paper we denote the position vector by $\br=(\bx,z)$, whose  horizontal component is $\bx=(x,y)$ and $z$ is height. Gravitational acceleration, $-g\zhat$, is constant with $g=274$~m\,s$^{-2}$. We  use a  plane-parallel geometry as an approximation to a local patch on the Sun: this is valid for wavelengths much less than the solar radius. We denote by  $\los$ the unit vector pointing toward the observer, which depends on the position on the solar disk but is assumed to be independent of position over the small solar region of interest. The line-of-sight vector $\los=(\los_{\rm h},\ell_z)$ decomposes into a horizontal component $\los_{\rm h}$ and a vertical component $\ell_z$.

The velocity field has two components: the propagating part, $\bbv(\br,t)$, and the underlying steady horizontal flow, $\bu(\br)$. As mentioned in the introduction and in order to obtain a simple problem, we choose both $\bu$ and $\bbv$ to be irrotational at all times. We introduce the wave velocity potential, $\Theta$, such that
\begin{equation}
\label{potential}
\bbv=\bnabla\Theta.
\end{equation}
The main advantage of this setting is that the problem of surface-gravity wave propagation through flows reduces to a two-dimensional one.

As is appropriate for solar waves, we consider small-amplitude (linear) waves. 
The linearized hydrodynamic equations we wish to solve are as follows. In the bulk $z<0$, the momentum and continuity equations are
\begin{eqnarray}
\label{mom-eq}
-\ii \omega \Theta + \bu\cdot\bnabla\Theta  &=& - p/\overline{\rho} - \Gamma[\Theta] ,\\ 
\bnabla^2 \Theta &=& 0, 
\end{eqnarray}
where $p$ is the pressure perturbation,  $\overline{\rho}$ is
the density, and $\Gamma$ is a phenomenological frequency-dependent damping operator \citep[see][]{gizon2002}. The above equations and the ones that follow apply to quantities that were Fourier-transformed in time, with $\omega$  the angular frequency \citep[we use the Fourier conventions of][]{gizon2002}.   At the free surface surface $z =0$, the linearized dynamical and kinematic boundary conditions \citep[e.g.,][]{whitham1974} are
\begin{eqnarray}
p/\overline{\rho} -  g \eta &=& \Pi/\overline{\rho} \qquad z=0 , \\ 
-\ii \omega \eta + \bu\cdot\bnabla\eta &=& \partial_z \Theta  \qquad z=0 , \label{surf-eq}
\end{eqnarray}
where $\eta$ is the elevation of the fluid's surface, and $\Pi$ is a stochastic pressure source on the surface that generates the waves \citep{gizon2002}.

In many cases, observations of solar oscillations are Doppler velocity observations \citep[e.g.][]{scherrer1995}. Let us introduce the line-of-sight component of the wave velocity at the surface and at time $t$, $\phi(\bx,t)$, given by
\begin{equation}
\phi(\bx,t) =  \los\cdot\bbv(\bx,z=0,t)  .
\end{equation}
The actual measurements involve convolution of $\phi$ with the point-spread function of the telescope. In addition, the f-mode data analysis includes filtering the observations to select the f-mode ridge in Fourier space \citep{duvall2000}.  We denote the resulting data cube by $\psi(\bx,t)$. The combined effects of the point-spread function and the data filtering can be described by the convolution of a function $F(\bx,t)$ with $\phi(\bx,t)$:
\begin{equation}
\label{psi-phi}
\psi(\bx,t) = (F \star \phi)(\bx,t)  .
\end{equation}
In Fourier space, the filtered data $\psi(\bk,\omega)$, where $\bk$ is the horizontal wave vector, is given by a multiplication of the filter $F(\bk,\omega)$ by $\phi(\bk,\omega)$, 
\begin{equation}
\psi(\bk,\omega) = F(\bk,\omega) \phi(\bk,\omega)  .
\end{equation}
As in \citet{gizon2002}, a mode-mass correction is also included in $F$ to simulate the response of a solar-like stratified atmosphere.  Note that we use the same symbol for a function and its Fourier transform: the arguments are used to distinguish the two.

We are first interested in the filtered oscillation signal $\psi^0$ that would be observed in the absence of any flows (zero-order terms are denoted with a superscript 0).  Eliminating the pressure and surface elevation terms, the surface boundary conditions at $z=0$ reduce to 
\begin{equation}
\label{source}
(\partial_z -\kappa)\Theta^0 = \frac{\ii\omega}{\overline{\rho} g} \Pi\equiv S^0 , 
\end{equation}
where  $\kappa(\omega) = (\omega + \ii \gamma)\omega/g$ is the complex
wavenumber at resonance. The zero-order damping operator is such that
$\Gamma^0[\Theta^0] = \gamma \Theta^0$ with the damping frequency
$\gamma(\omega)/2\pi = |\omega/\omega_*|^{4.4} \times  100\, \mu{\rm
  Hz}$, and $\omega_*/2\pi=3$~mHz is the central frequency \citep{gizon2002}.

The solution of equation~(\ref{source}) can be found by introducing
a Green's function. The final result for $\psi^0$ (on the surface) is then
\begin{equation}
\label{psi0}
\psi^0(\bx,\omega)  =  F\left[ (\ell_{\rm h}\cdot\bnabla_{\rm h} + \ell_z\kappa) \Theta^0   \right],
\end{equation}
where $\bnabla_{\rm h}$ is the horizontal gradient. The zero-order surface velocity potential is given by
\begin{equation}
\Theta^0(\bx, z=0, \omega)  =  2\pi\doubint G(\bx-\bx',\omega)S^0(\bx',\omega) \id^2\bx'  , 
\end{equation}
with 
\begin{equation}
\label{green}
G(\bx,\omega) = \frac{1}{(2\pi)^3} \doubint \frac{e^{\ii\sbk\cdot\sbx}}{k-\kappa(\omega)} \id^2\bk,
\end{equation}
where $k=\|\bk\|$ is the horizontal wavenumber. In summary,
equations~(\ref{source})-(\ref{green}) give the observed zero-order filtered wave field at the
surface for f modes propagating through the background model with no flows
present.


\section{Travel-time shift due to a uniform flow}
\label{sec:uniform}
Here we study the effect of a horizontal, spatially-uniform steady flow,
\begin{equation}
\bu=u_x \hat{\bx} + u_y \hat{\by} = {\rm const} .
\end{equation}
From equations~(\ref{mom-eq})-(\ref{surf-eq}) we see that the effect of the flow is to transform $\omega$ into $\omega - \bk\cdot\bu$.  The Doppler shift $\bk\cdot\bu$ is the result of a simple Galilean transformation $\bx\rightarrow\bx-\bu t$. This transformation applies to
all quantities in the problem except the filter $F$, which is applied independently of the flow. The filtered data cube can thus be expressed in terms of the Doppler velocity in the absence of a flow, $\phi^0$, according to
\begin{equation}
\label{psi-flow}
\psi(\bk,\omega) = F(\bk,\omega) \phi^0(\bk,\omega-\bk\cdot\bu) .
\end{equation}
     
\begin{figure}[t]
  \includegraphics[ width=0.9\linewidth ]{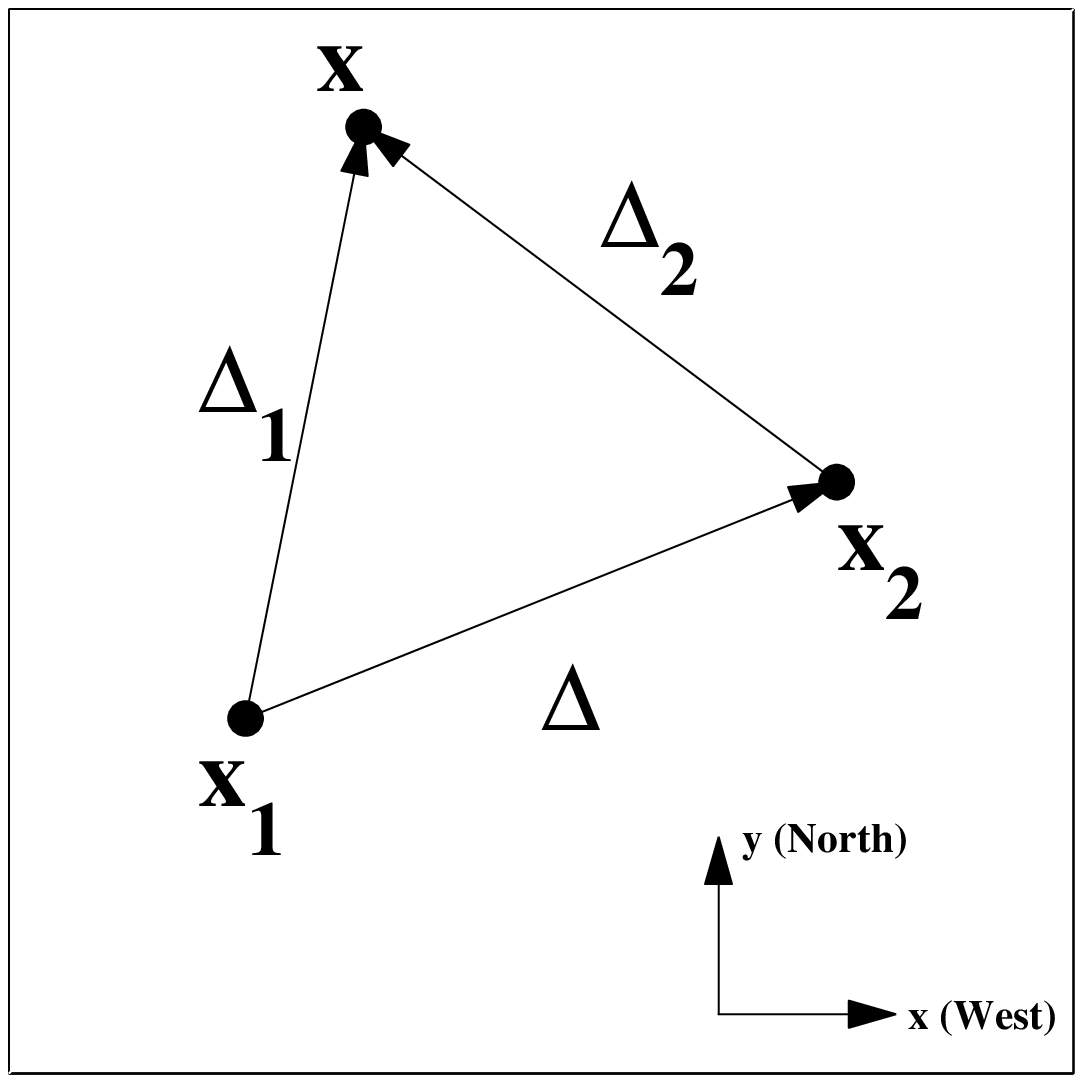}
  \centering
  \caption{Schematic of the geometry of the plane-parallel patch of the Sun we consider.   Travel times are measured using the wave signal at the two photospheric
  observation points  $\1$ and $\2$. The distance between these points
  is $\Delta=\|\bDelta\|$. The vectors given by $\bDelta_1$ and
  $\bDelta_2$ appear in the expression for the sensitivity kernels (see Appendix~\ref{app:curlyc}). }
\label{fig:geometry}
\end{figure}

The fundamental computation in time-distance helioseismology is the cross-covariance of the oscillation signal, measured for two different points $\1$ and $\2$ on the solar surface, 
\begin{equation}
C(\1,\2,\omega)= \frac{2\pi}{T} \psi^*(\1,\omega) \psi(\2,\omega)  ,
\end{equation}
where the star denotes complex conjugation and $T$ is the observation duration. Because of horizontal translation invariance (uniform flow), the expectation value of the cross-covariance is the inverse Fourier transform of the power spectrum \citep[e.g.][]{gizon2004}:
\begin{equation}
\label{cross-cov}
C(\1,\2,\omega) = \doubint P(\bk,\omega) e^{\ii\sbk\cdot\sbDelta} \, \id^2\bk ,
\end{equation}
where $\bDelta=\2-\1$ and 
\begin{equation}
\label{pow}
P(\bk,\omega) = E[|\psi(\bk,\omega)|^2]
\end{equation}
is the expectation value of the power spectrum of oscillations. The computation of $P$ requires the knowledge of the source covariance function $m_s(\omega)$; here we choose the same function as in \citet{gizon2002}.
A general configuration of the observation points and notation used throughout the paper is depicted in  Figure~\ref{fig:geometry}.

\begin{figure*}[t]
   \includegraphics[ width=0.95\linewidth]{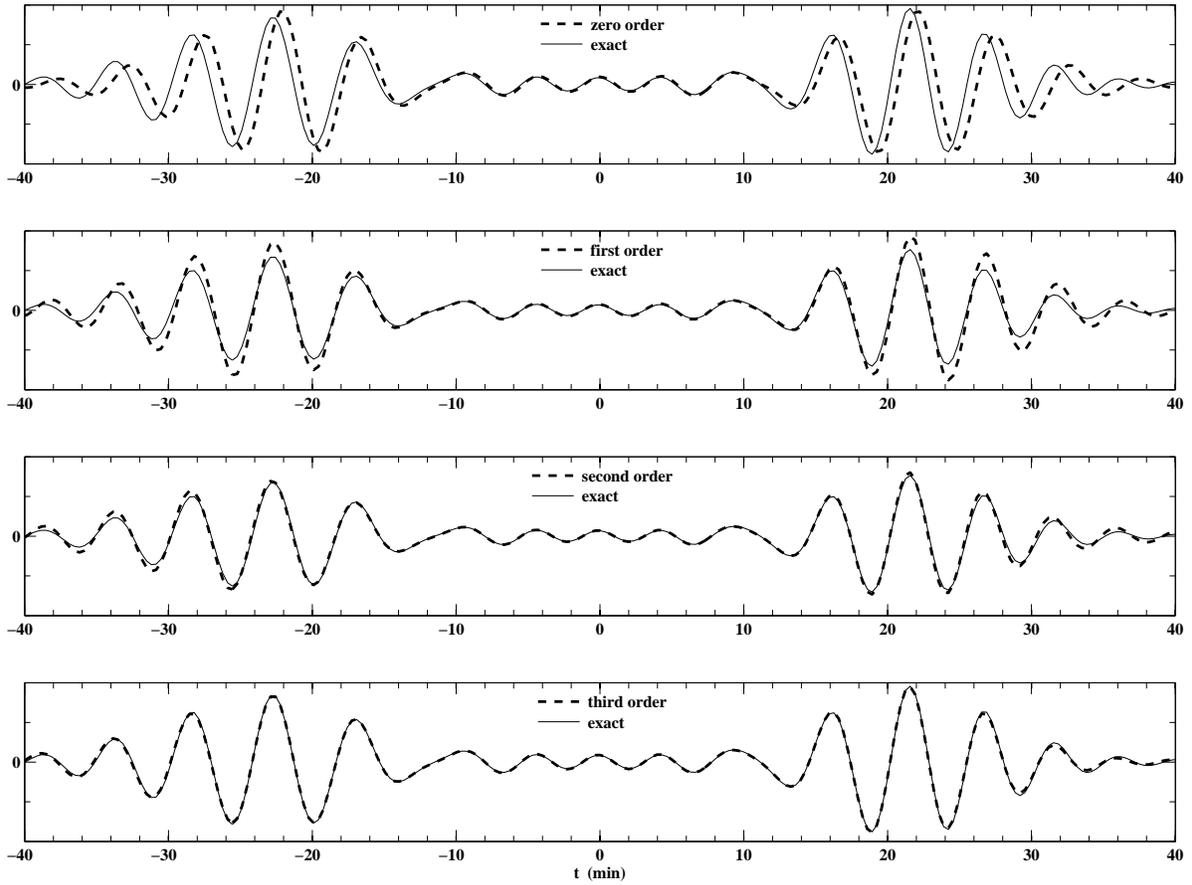}
\centering
   \caption{Temporal cross-covariance functions $C(\1,\2,t)$ in the uniform flow model for $\bu= 400$~m\,s$^{-1} \hat{\sbDelta}$ and  $\Delta=10$~Mm. The thin solid lines are the exact cross-covariances (repeated in each panel), and the thick dashed lines correspond to  the approximate cross-covariances, truncated at the order indicated in the legend of each panel. }
 \label{fig:cross_corr}
\end{figure*}

\begin{figure}[t]
  \includegraphics[ width=\linewidth]{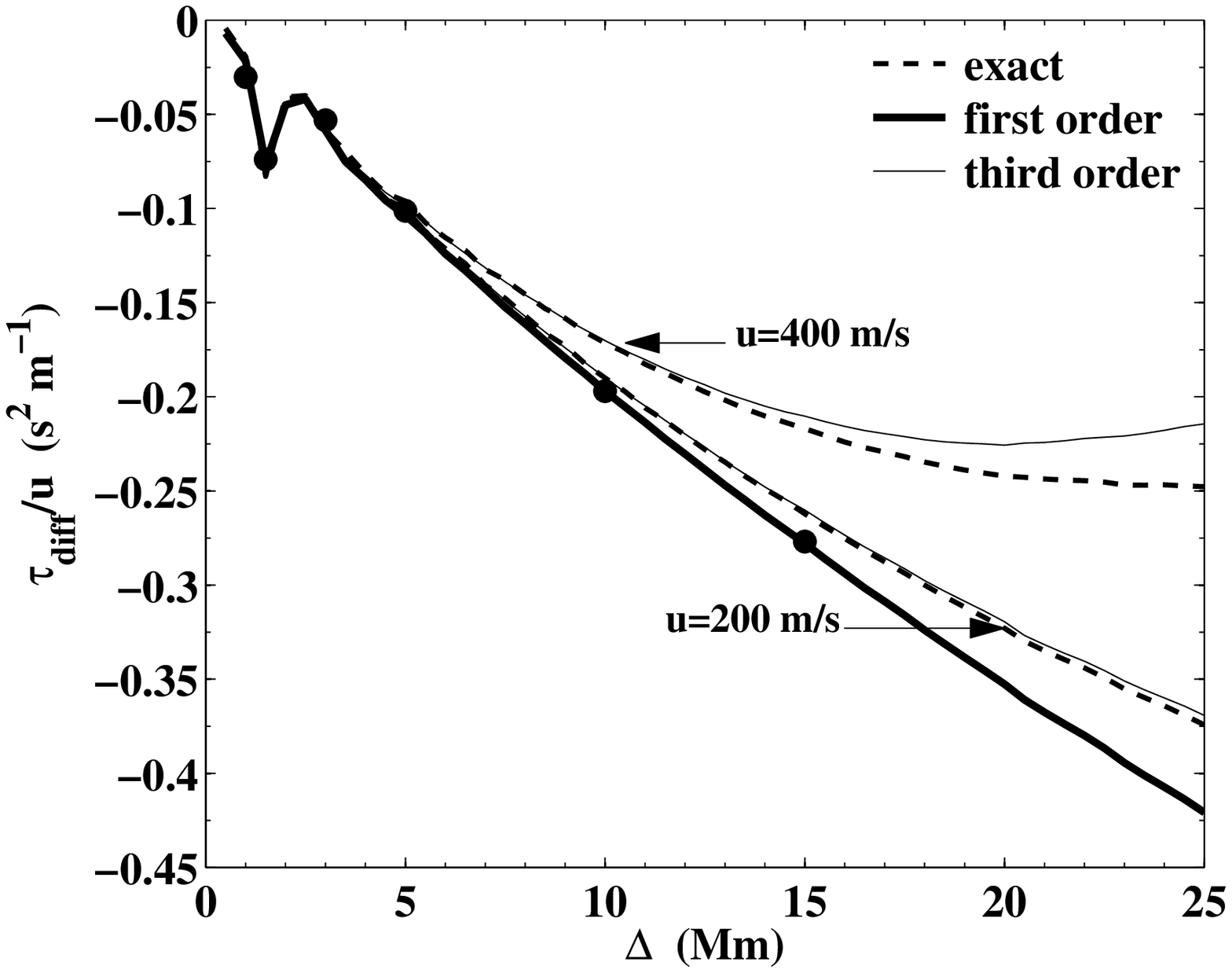}
  \centering
  \caption{Scaled travel-time differences in the uniform flow model as a
  function of separation distance $\Delta$ for two different flow speeds, where $\bu=u\hat{\sbDelta}$. The thick solid line is the
    linear approximation (which is independent of the flow when plotted as
    $\tau/u$). The upper and lower dashed lines correspond to the exact
    solutions for $u=400$~m\,s$^{-1}$ and $u=200$~m\,s$^{-1}$ respectively. The thin solid
    lines are the third-order approximation for each $u$. The big dots denote the travel-time differences derived from integration of the 
    kernel ${\rm K}^{\rm diff}_x$  discussed in detail in \S~\ref{sec:ptp}.}
\label{fig:tt_v_d}
\end{figure}

We now consider two strategies for the computation of the cross-covariance $C$ in the case of a uniform flow. The first is to compute  equations~(\ref{cross-cov}) and (\ref{pow}) numerically, to obtain what we call the exact solution.
The second is to compute successive approximations to $C$ using a Taylor expansion  in $\bu$ of the power spectrum $P$.  For a constant flow, we have 
\begin{eqnarray}
\label{expansion}
\nonumber
P (\bk,\omega) & = &  P^0 (\bk,\omega) -\bk\cdot\bu |F|^2\partial_{\omega}\left[|\phi^0|^2\right] \\ \nonumber
&+&
\frac{1}{2}(\bk\cdot\bu)^2|F|^2\partial_{\omega}^2\left[|\phi^0|^2\right] \\
&
- &\frac{1}{6}(\bk\cdot\bu)^3|F|^2\partial_{\omega}^3\left[|\phi^0|^2\right]
+\dots ,
\end{eqnarray}
where $P^0(\bk,\omega)=E[|\psi^0|^2]$ is the expected power spectrum when $\bu=0$. We may truncate this expansion after each successive term to study the zero-, first-, second-, and third-order cross-covariance (terms in $u^0$, $u^1$, $u^2$ and $u^3$ respectively).  

Figure~\ref{fig:cross_corr} shows the cross-covariance function at each level of approximation of the Taylor expansion for a uniform flow. Convergence to the exact covariance is observed as higher-order terms are retained. Notice that the phase shift is largely accomplished by the first-order term, while the amplitude correction is largely accomplished by the second-order term.

\begin{figure}[t]
   \includegraphics[ width=\linewidth]{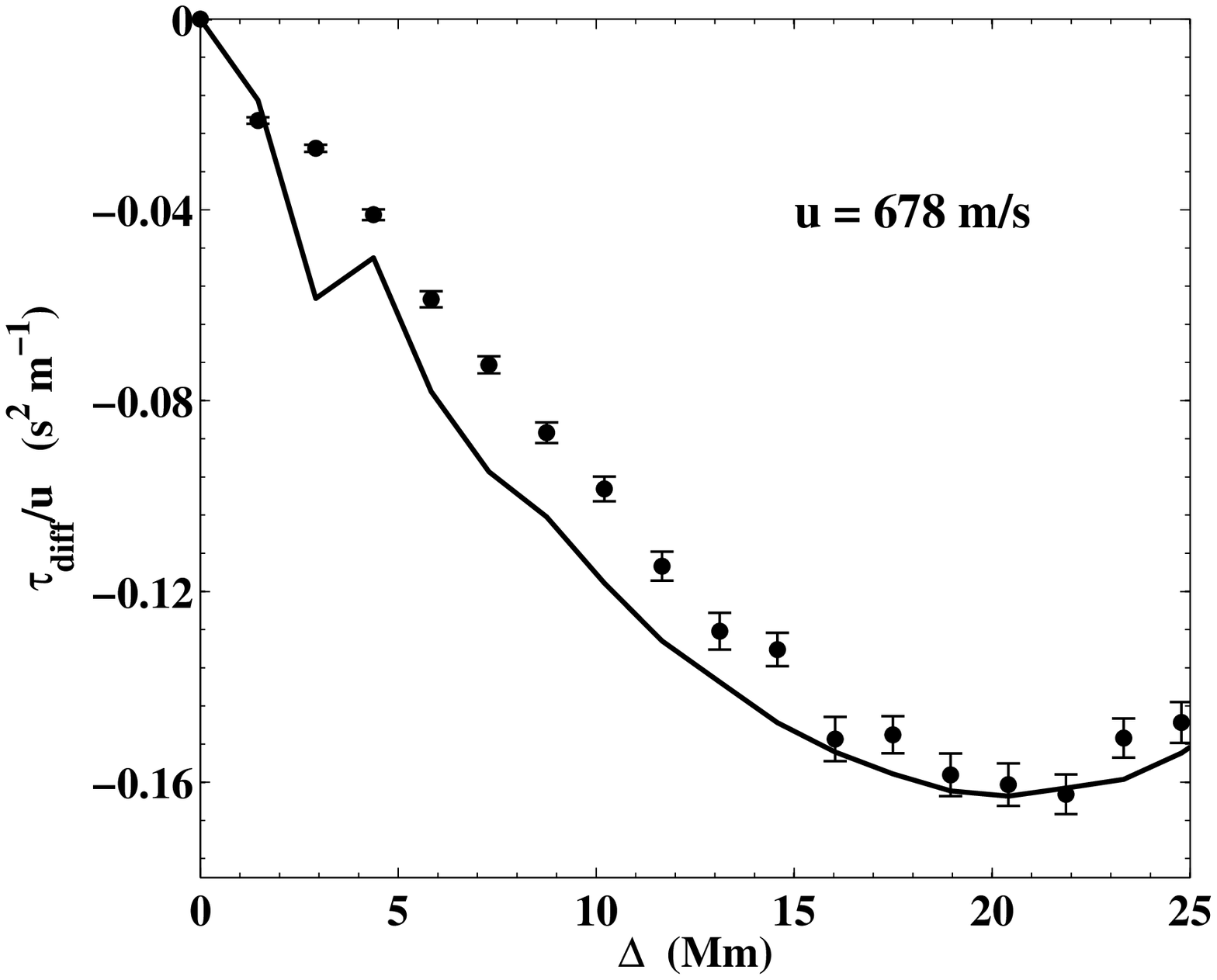}
\centering
   \caption{Comparison of travel-time differences from data and from the exact
     solution for a uniform flow. The circles are the travel-time differences measured from quiet sun MDI Dopplergrams tracked at the constant rate of  678~m\,s$^{-1}$. The solid line denotes the travel-time differences computed from the exact model of \S~\ref{sec:uniform} for a constant flow of 678~m\,s$^{-1}$. The one standard deviation error bars of the data are also shown.}
 \label{fig:tt_data_exact}
\end{figure}

\begin{figure}[t]
   \includegraphics[ width=\linewidth]{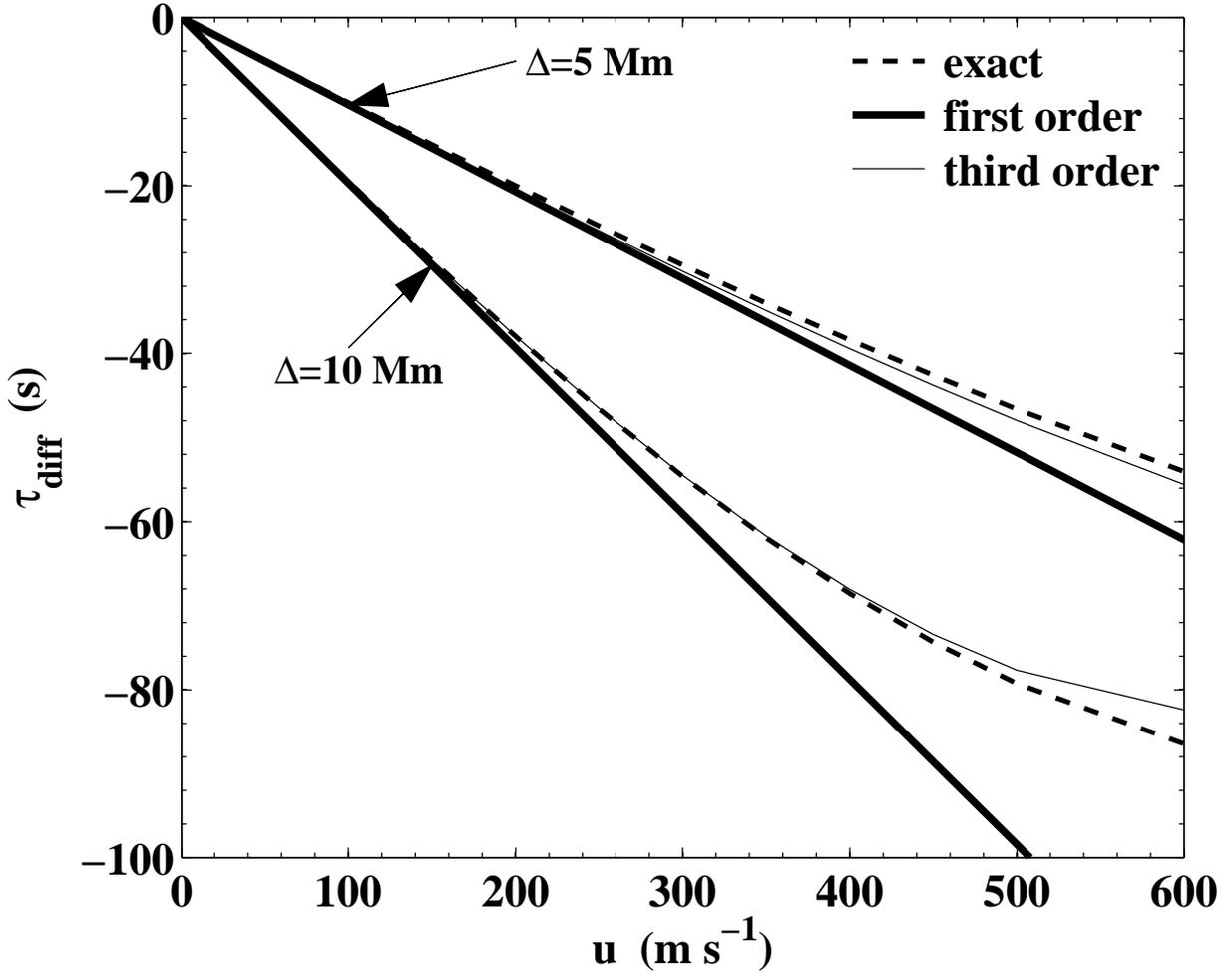}
\centering
   \caption{Travel-time differences as a function of uniform flow speed $u$ for two
     different distances $\Delta$. For each set of curves,   the dashed line is the exact solution, the thick solid
     line is the first-order approximation, and the thin solid line is the third-order
     approximation. The top set (top three curves) of
     travel-time differences is for $\Delta=5$~Mm, while the bottom set (bottom three curves) is
     for $\Delta=10$~Mm.}
 \label{fig:tt_v_u}
\end{figure}

\begin{figure}[t]
   \includegraphics[ width=\linewidth]{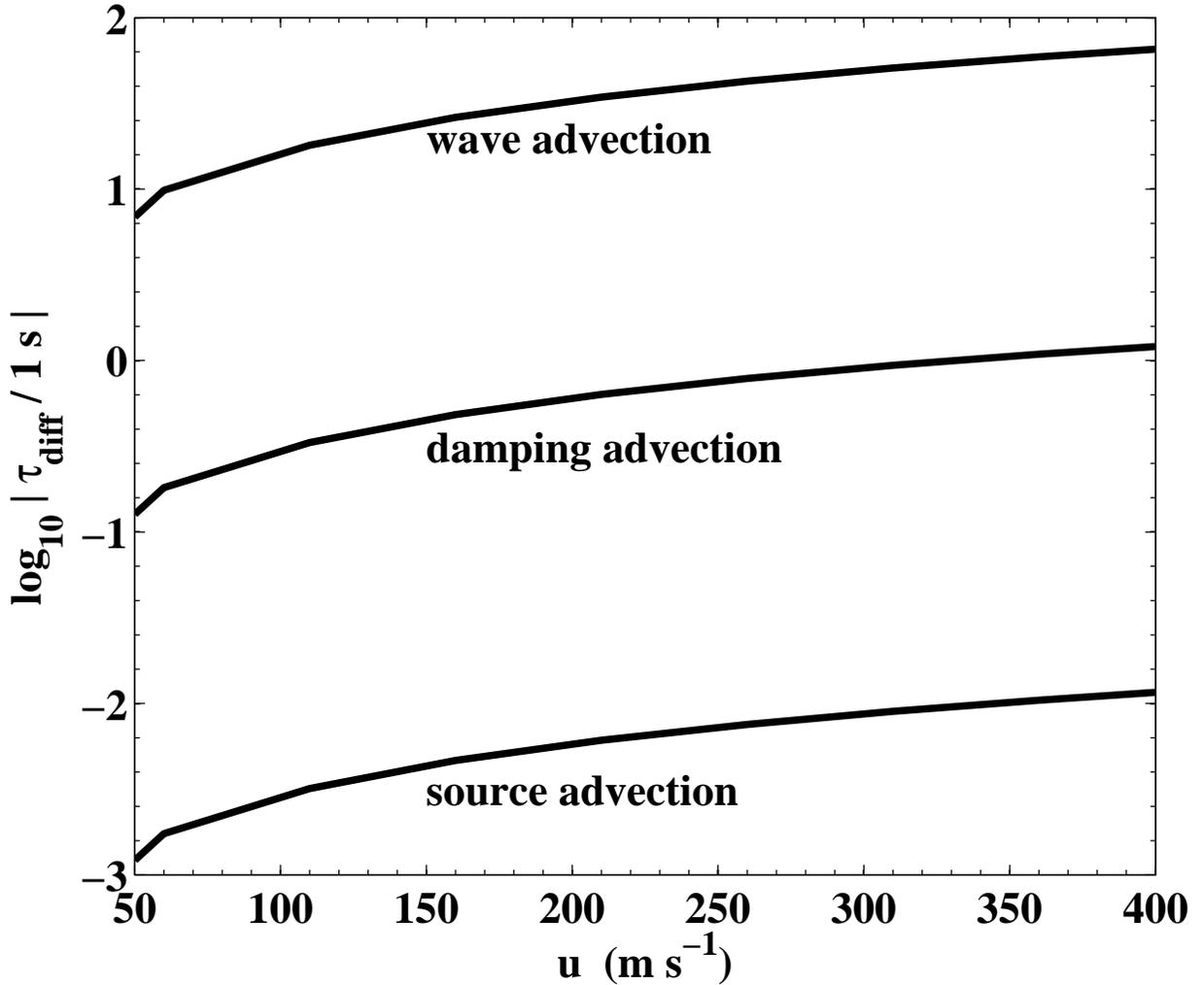}
\centering
   \caption{First-order travel-time difference contributions from the three separate advection
     terms of equation~(\ref{deltac_constant}) for a uniform flow. The $y$ coordinate scale is the $\log$ of the absolute value of the travel-time difference per second. In this case,
     $\Delta=10$~Mm and $\bu=u\hat{\sbDelta}$. }
 \label{fig:tt_3terms}
\end{figure}

\begin{figure*}[t]
   \includegraphics[ width=\linewidth]{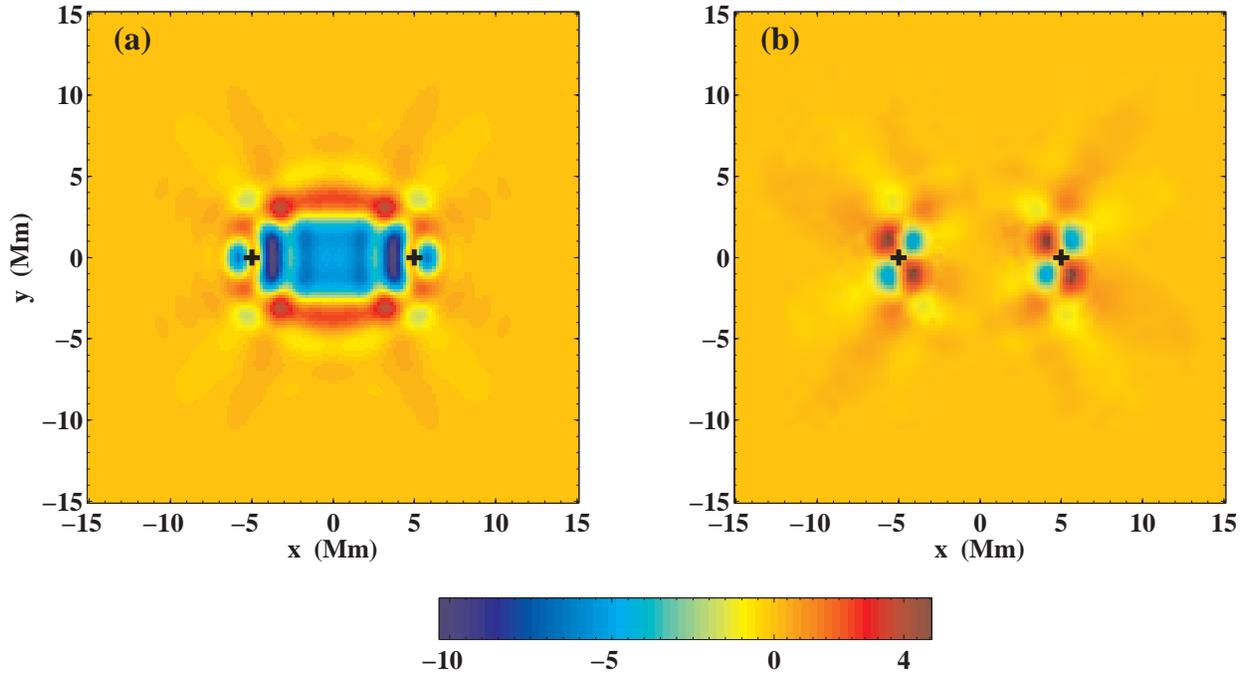}
\centering
    \caption{F-mode travel-time difference sensitivity kernel for flows as a function of horizontal position $\bx=(x,y)$.   Panel~(a) shows ${\rm K}_x^{\rm diff}$ and panel~(b) shows ${\rm K}_y^{\rm diff}$, with $\bDelta=10$~Mm $\hat{\bx}$ in both cases. The crosses denote the two observation points at $\1=(-5,0)$~Mm and $\2=(5,0)$~Mm. The units of the kernel are
      s\,Mm$^{-2}$ (km/s)$^{-1}$.}
  \label{fig:kerns10}
\end{figure*}

In this paper, we are mainly interested in studying travel-time differences, which are obtained by finding the time $\tau_+(\1,\2)$ it takes for waves travelling from $\1$ to $\2$, and subtracting the time $\tau_-(\2,\1)$ for those waves going from $\2$ to $\1$. In other words, 
\begin{equation}
\tau_{\rm diff}(\1,\2)=\tau_+(\1,\2)-\tau_-(\2,\1).
\end{equation}
This quantity gives information about the underlying flows. Following \citet{gizon2004},  the travel-time difference can be measured from the cross-covariance function according to
\begin{equation}
\label{tau-w}
\tau_{\rm diff}(\1,\2)=\frac{2\,{\rm Im}\int_0^{\infty} \omega\, C^0(\bDelta,\omega)\,\Delta C(\1,\2,\omega) \, \id\omega  }{\int_0^{\infty}\omega'^2 |C^0(\bDelta,\omega')|^2 \, \id\omega'},
\end{equation}
where $C^0(\bDelta,\omega) = C^0(\1,\2,\omega) $ is the covariance function in the absence of a flow and $\Delta C=C-C^0$ is the change in the cross-covariance due to the flow. The operator Im takes the imaginary part of the expression. 

Using the zero-, first-, and third-order cross-covariances, as well as the exact cross-covariance, we now  study travel-time differences using equation~(\ref{tau-w}).  This will provide us with a means to test the first-order Born approximation which is developed in \S~\ref{sec:born}, as well as to quantify higher-order terms. Note that the second-order term does not affect travel-time differences because it only introduces perturbations to the cross-covariance that are symmetric in the time lag.

We first study the travel-time differences for various separation distances $\Delta$ and flow speeds $u$. Figure~\ref{fig:tt_v_d} shows the exact and approximate scaled travel-time differences $\tau_{\rm diff} / u$ versus $\Delta$ computed at fixed $\bu = u \hat{\bDelta}$ for $u=200$~m\,s$^{-1}$ and $u=400$~m\,s$^{-1}$ ($\hat{\bDelta}=\bDelta/\Delta$). We find that the first-order approximation is within 10\% of the exact value for $u=200$~m\,s$^{-1}$ and $\Delta<20$~Mm, while the error is only about 1\% when terms up to $u^3$ are kept.  For a flow that is twice as large, $u=400$~m\,s$^{-1}$, non-linear effects are more important. The functional dependence at large $\Delta$ is partly due to mode damping, as will be shown later in \S~\ref{sec:ray}.

Next we compare this result with observed travel times.  We used 4~hours of quiet Sun MDI Dopplergrams tracked at a constant rate of $678$~m\,s$^{-1}$ with respect to the local surface rotation rate in order to mimic the effect of a constant flow. Travel times are  then measured using equation~(\ref{tau-w}).  Figure~\ref{fig:tt_data_exact} shows the result. We find a good agreement between the exact model  and the observations. The agreement is not, however, perfect at short distances; this is likely to be due to a mismatch between the model and the solar f-mode power spectra.

Figure~\ref{fig:tt_v_u} shows the variations of $\tau_{\rm diff}$ as a function of $u$ for fixed distances $\Delta=5$~Mm and $\Delta=10$~Mm. The first-order approximation is within 10~\% of the exact solution for flows with amplitudes less than about 400~m\,s$^{-1}$, although for larger $\Delta$ the agreement becomes worse. However, these results demonstrate that the first-order approximation (and most likely the kernels given in \S~\ref{sec:generalflow}) is appropriate for the study of supergranular flows. 

Let us turn to the question of whether advection of the waves by the flow is the only important effect. To first order in $u$, the change in the cross-covariance in the case $\los=\hat{\bz}$ has three distinct components: 
\begin{eqnarray}
\label{deltac_constant}
\nonumber
\Delta C &\simeq& -(2\pi)^6\doubint\id^2\bk\, e^{\ii\sbk\cdot\sbDelta}\, (\bk\cdot\bu)\, |F|^2\\\nonumber
&\times&\left[m_s|\kappa|^2 \partial_\omega(|G|^2) + (\partial_\omega m_s) |\kappa|^2|G|^2\right. \\
&+&\left. m_s|G|^2 \partial_\omega(|\kappa|^2)\right] .
\end{eqnarray}
The three terms  in brackets in this equation correspond to the  effect of the flow on the waves, on the sources, and on the damping. We can thus determine the relative contribution that each of these physical effects separately has on the first-order travel-time differences. This is shown in Figure~\ref{fig:tt_3terms}. The advection of the waves contributes by far the most to the travel-time difference, making up about 98\% of the total value. We conclude that, for practical applications, we can ignore the effects of the flow on the damping and source functions.


\section{Spatially-varying horizontal flow}
\label{sec:generalflow}
Now we relax the assumption of a constant flow  and consider a spatially-varying, steady, horizontal, small-amplitude flow  $\bu(\bx)$. The general study of the interaction of surface waves with flows involves a number of complications, and we do not attempt to solve the problem in its full generality; instead we require  the flow vorticity be zero.  Due to the small amplitude  of the flow, the problem may be solved in the first Born approximation, which we carry out in \S~\ref{sec:born} and  will lead to the sensitivity kernels of \S~\ref{sec:ptp}.

\subsection{First-order Born approximation}
\label{sec:born}
Here we solve the linearized set of hydrodynamic equations introduced in \S~\ref{sec:surfacewaves} by applying the  first Born approximation.  We expand each wave quantity $q$ into a zero-order term, $q^0$, and a perturbation term due to the flow, $\delta q$. We then insert the expanded wave quantities into equations (\ref{mom-eq})-(\ref{surf-eq}) and retain only terms which are of first order.   In the bulk, the perturbed velocity potential $\delta\Theta$ satisfies the three-dimensional Laplace's equation,
\begin{equation}
\bnabla^2  \delta\Theta =0  .
\end{equation}
Eliminating the pressure and the surface elevation, the surface boundary conditions at $z=0$ reduce to
\begin{equation}
(\partial_z -\kappa)\delta\Theta = \ii \bu\cdot\bnabla \left[(\partial_\omega\kappa)\Theta^0 + \partial_\omega S^0\right] \equiv \delta S .
\label{eq.eqds}
\end{equation} 
In doing so, we have used the approximations $\delta\Pi\approx\ii \bu\cdot\bnabla \partial_\omega\Pi$ and $\delta \Gamma \Theta^0 \approx \ii  (\partial_\omega\gamma) \bu\cdot\bnabla \Theta^0$, which describe the perturbations to the wave sources and the wave damping respectively.
We note that $\delta \Theta$ satisfies the same equations as $\Theta^0$, except for a different source function $\delta S$  on the right-hand side of the surface boundary condition, equation~(\ref{eq.eqds}). This is the first Born approximation, an equivalent-source description of wave scattering. 

Using the same Green's function as in \S~\ref{sec:uniform}, the first-order surface velocity potential is
\begin{equation}
\delta\Theta(\bx, z=0, \omega)  =  2\pi\doubint G(\bx-\bx',\omega)\delta S(\bx',\omega) \id^2\bx'  ,
\end{equation}
from which we deduce the perturbation to the filtered data,
\begin{equation}
\label{dpsi}
\delta\psi(\bx,\omega)  = F\left[ (\ell_{\rm h}\cdot\bnabla_{\rm h} + \ell_z\kappa) \delta \Theta  + \ii \ell_z(\partial_\omega\kappa) \bu\cdot\bnabla \Theta^0\right].
\end{equation}


\subsection{Linear sensitivity kernels for flows}
\label{sec:ptp}

The first-order perturbation to the cross-covariance is obtained by approximating $\Delta C$ in equation~(\ref{tau-w}) according to 
\begin{eqnarray}
\label{deltac}
\nonumber
\Delta C(\1,\2,\omega)& \simeq & E\left[\delta\psi^*(\1,\omega)\psi^0(\2,\omega)\right.\\
&+&\left.\psi^{0*}(\1,\omega)\delta\psi(\2,\omega)\right] ,
\end{eqnarray}
where $\delta\psi$ is the  first-order perturbation to $\psi$. Using the expressions we have found for $\psi^0$ and $\delta\psi$,  equation~(\ref{deltac}) can be put into the form 
\begin{equation}
\label{curlyc}
\Delta C(\1,\2,\omega) \simeq \doubint \cC(\1,\2,\omega;\bx)\cdot\bu(\bx)\,\id^2\bx,
\end{equation}
where the function $\cC$, explicitly written in Appendix~\ref{app:curlyc},  gives the linear sensitivity of the cross-covariance to a spatially-varying flow. Inserting equation~(\ref{curlyc}) into equation~(\ref{tau-w}) gives
\begin{equation}
\label{tau-k} \tau_{\rm diff}(\1,\2) = \doubint\bK^{\rm diff}(\1,\2;\bx)\cdot\bu(\bx)\,\id^2\bx ,
\end{equation}
where the function $\bK^{\rm diff}$ is the two-dimensional sensitivity kernel, defined by
\begin{equation}
\label{K}
\bK^{\rm diff} = \frac{2 \; {\rm Im}\int_0^{\infty} \omega
  C^0(\bDelta,\omega) \cC(\1,\2,\omega;\bx)\,\id\omega}{\int_0^{\infty}\omega'^2 |C^0(\bDelta,\omega')|^2 \, \id\omega'}.
\end{equation}

\begin{figure*}[t]
   \includegraphics[ width=0.9\linewidth]{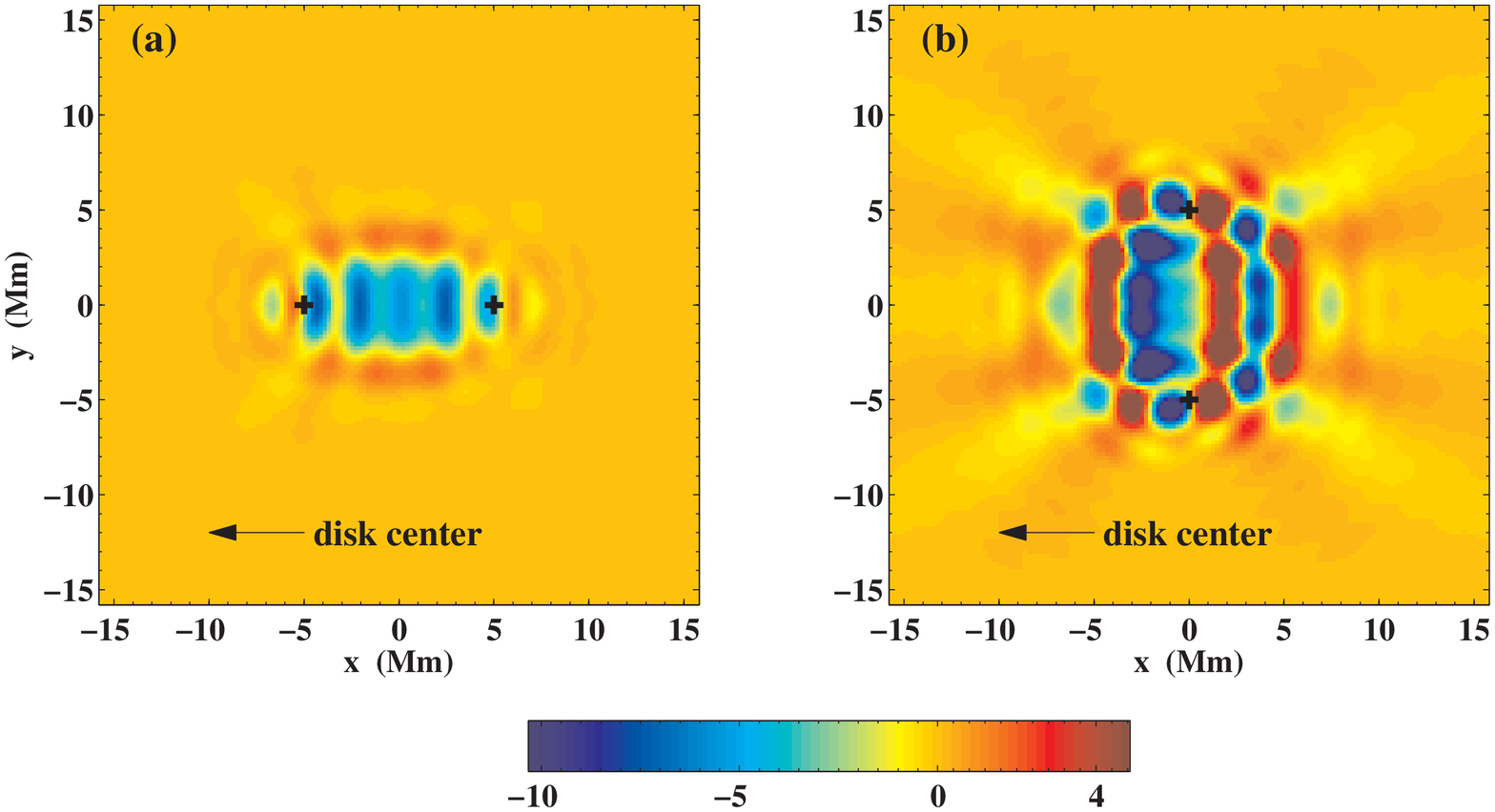}
\centering
    \caption{F-mode travel-time sensitivity kernels computed away from disk center, at $\Phi=\pi/3$,
      $L=0$ (on the equator towards western limb). The definition of the angles is given by equation~(\ref{los}). The two observation points are denoted
      by the crosses. (a) ${\rm K}_x^{\rm diff}$ for $\1=(-5,0)$~Mm, $\2=(5,0)$~Mm. (b) ${\rm K}_y^{\rm diff}$ for the case when
      $\1=(0,-5)$~Mm, $\2=(0,5)$~Mm. The direction towards the disk center is annotated. The units are s\,Mm$^{-2}$ (km/s)$^{-1}$.}
  \label{fig:kerns10_los}
\end{figure*}

Figure~\ref{fig:kerns10} shows example kernels for the linear sensitivity of the travel-time difference to small-amplitude flows, for a separation of 10~Mm between the observation points and for a vertical line of sight $\los=\hat{\bz}$. A short discussion of the numerical computation of kernels is given in Appendix~\ref{sec:computation}. The kernel in Figure~\ref{fig:kerns10}a, ${\rm K}_x^{\rm diff}$, gives the sensitivity of the travel times to $u_x$ and the
kernel in Figure~\ref{fig:kerns10}b, ${\rm K}_y^{\rm diff}$, gives the sensitivity to $u_y$.  For reasons of symmetry (the ${\rm K}_y^{\rm diff}$ kernel is antisymmetric with respect to the lines $x=0$ and $y=0$) the total integral of ${\rm K}_y^{\rm diff}$ is zero.

We  observe elliptical and hyperbolic features in the kernels, similar to what was seen in the kernels for the damping rate and the source strength perturbations derived by \citet{gizon2002}. The elliptical features have been called Fresnel zones in geophysics \citep[e.g.,][]{tong1998}. The first Fresnel zone bounds the large area between the observation points in Figure~\ref{fig:kerns10} where the kernel is negative. That  ${\rm K}^{\rm diff}_x$ is negative in this region is intuitive: local flows in the $+\hat{\bx}$ direction should speed up waves traveling in this direction while slowing down waves going the opposite way, resulting in  negative travel-time differences.  The hyperbolae and ellipses, due to scattering of waves generated by distant excitation events, are discussed futher in \S~\ref{sec:farfield}.

It is important and instructive to check the Born kernels for consistency with the first-order approximation of $\tau_{\rm diff}/u$ calculated under the assumption of a constant flow in
 \S~\ref{sec:uniform}. It also enables us to verify the numerical computation of the kernels. If we again consider a uniform flow, $\bu=u\hat{\bx}$, then from equation~(\ref{tau-k}) it is possible to spatially integrate the kernels to obtain
\begin{equation}
\label{integrate_kerns}
\frac{\tau_{\rm diff}(\bDelta)}{u}=\doubint {\rm K}_{x}^{\rm diff}(\bDelta;\bx)\,\id^2\bx .
\end{equation}
We have computed kernels for several values of $\Delta$ and obtained the corresponding $\tau_{\rm diff}/u$. The results are shown as large dots in Figure~\ref{fig:tt_v_d}. The total integral of ${\rm K}_x^{\rm diff}$ is within $0.1$\% of the value expected from the first-order approximation, giving us confidence in the numerics.


\subsection{Effect of the line-of-sight projection}
\label{sec:lineofsight}

The kernels in Figure~\ref{fig:kerns10} were computed with the line-of-sight vector $\los=\hat{\bz}$, i.e., at solar disk center. We would like to obtain kernels at any general point on the solar disk. The variation in the line-of-sight vector across the solar disk is incorporated in our model and appears explicitly in the expression for the kernels (see Appendix~\ref{app:curlyc}). Foreshortening in the center-to-limb direction (the point-spread function effectively depends on the direction of $\bk$) is another important systematic effect, which has been ignored for the sake of simplicity.

A region on the Sun around the solar longitude measured from disk center, $\Phi$, and latitude, $L$, has a local line-of-sight vector approximately given by 
\begin{equation}
\label{los}
\los=-\sin\Phi\,\hat{\bx}-\sin L \cos\Phi\,\hat{\by}+\cos L \cos\Phi\,\hat{\bz} .
\end{equation}
Two examples of kernels computed away from disk center are given in Figure~\ref{fig:kerns10_los} for coordinates $\Phi=\pi/3$ and $L=0$ (on the equator towards the
western limb). The kernel in Figure~\ref{fig:kerns10_los}a is ${\rm K}_x^{\rm diff}$, whereas the kernel in Figure~\ref{fig:kerns10_los}b is ${\rm K}_y^{\rm diff}$, whose observation points lie on the same meridian ($\hat{\bDelta}=\hat{\by}$). If the line-of-sight effect did not matter, then the two kernels in Figure~\ref{fig:kerns10_los} and the kernel of Figure~\ref{fig:kerns10}a would all be identical after rotation of Figure~\ref{fig:kerns10_los}b. This is not the case, however. 

In particular, although these three kernels have the same total integral (see below), the kernel of Figure~\ref{fig:kerns10_los}b looks very different. This is because, in the case of Figure~\ref{fig:kerns10_los}b, waves that scatter along the line that connects the two observation points have a small component along the line of sight and have thus a small contribution to the cross-covariance function. In general, the flow sensitivity is reduced when the scattered wave at the observation points is travelling perpendicular to the center-to-limb direction.  For example, there is about $50$~\% less sensitivity along the ray path connecting the two observation points for the kernel of Figure~\ref{fig:kerns10_los}b compared to the kernel of Figure~\ref{fig:kerns10}a at disk center.

We have computed the total integral of the kernel as a function of position on the solar disk. We studied two cases with $L=0$ and varying $\Phi$ in the range from $-\pi/2$ to $\pi/2$: one in which the observation points lie on the equator and the other in which the observation points lie on the same meridian.   The resulting travel-time differences
 show very little, if any, variation across the disk. This means that large scale flows are practically insensitive to line-of-sight effects. However, judging by the fine structure of the kernels, the line-of-sight effect is important for small-scale flows.

\subsection{Source advection is unimportant}
\label{sec:adv}

In the uniform flow model it was possible to separate the effects of advection of the waves, of the wave sources, and of the wave damping to determine the contribution of each to the travel-time difference  (see eq.~[\ref{deltac_constant}] and Fig.~\ref{fig:tt_3terms}). We concluded that for uniform flows the wave advection is by far the dominant contribution to the travel time shifts. Is this also true for a spatially-varying flow? Using equation~(\ref{fullcurlyc}), it is possible to identify the contribution to the kernel due to source advection. At $\Delta=10$~Mm we find that the source advection contribution is about three orders of magnitude less than the contribution from the advection of waves by the flow.  This result is useful for future kernel calculations. Neglecting source advection is an excellent approximation. This approximation has been used in the past by, e.g., \citet{woodard2002}, \citet{birch2007}, and \citet{birch2007ring}, although not explicitly justified.

\subsection{Kernels depend on central frequency and bandwidth}
\label{sec:filtering}

\begin{figure}[t]
   \includegraphics[width=0.8\linewidth]{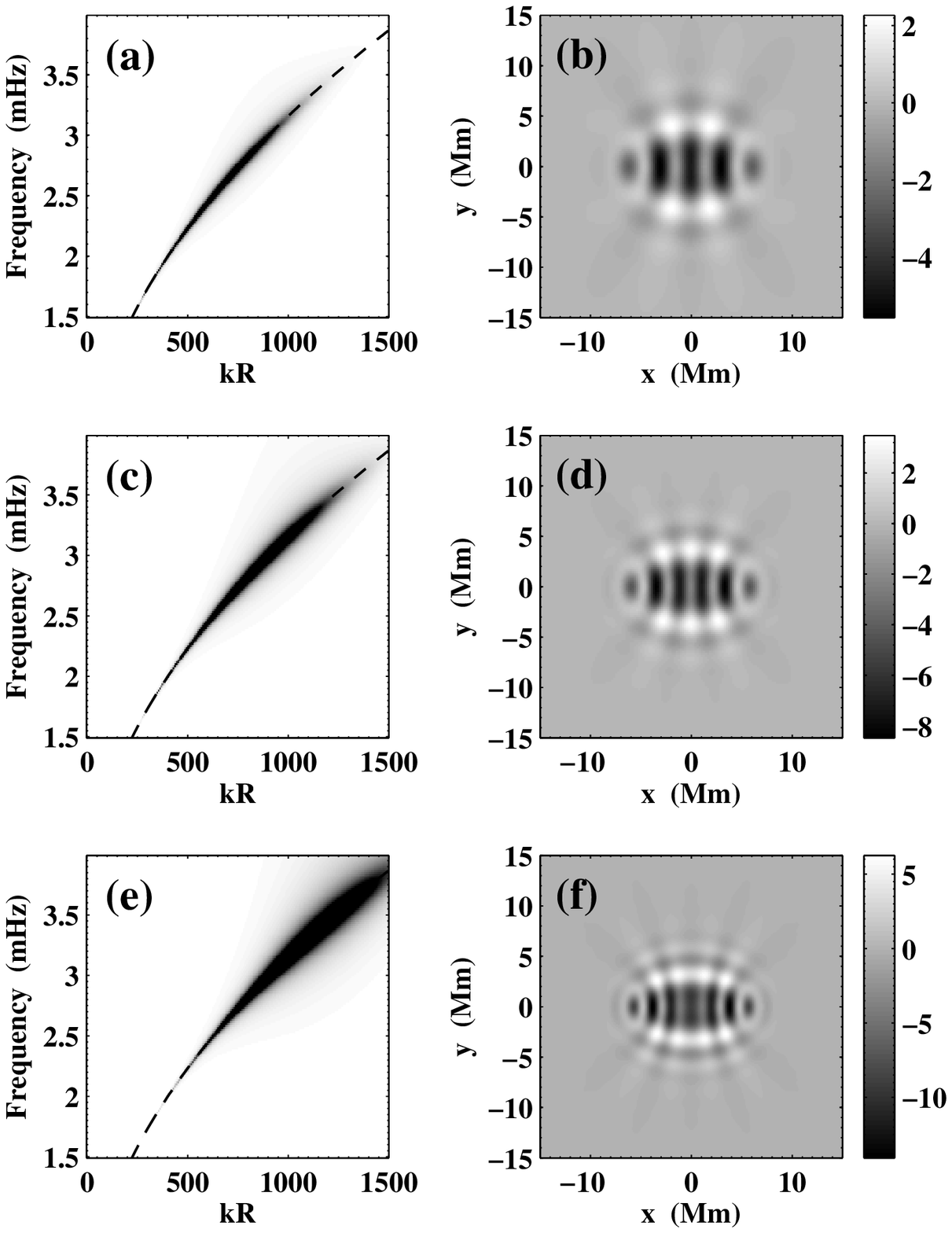}
\centering
\caption{Power spectra (\textit{a}), (\textit{c}), and (\textit{e}) and sensitivity kernels ${\rm K}^{\rm diff}_x$  (\textit{b}), (\textit{d}), and (\textit{f}) computed using the additional filter $G(k)$ defined in equation~(\ref{gfilt}). The filter used for the top row has a central wavenumber $k_0R=487$, for the middle row $k_0R=900$, and for the bottom row $k_0R=1322$. In each plot $\sigma R=348$, and $R=696$~Mm. For each kernel, $\Delta=10$~Mm, and the units are s\,Mm$^{-2}$ (km/s)$^{-1}$. The dispersion relation for f modes, $\omega^2=gk$, is given by the dashed lines in the power spectra. The gray scale of the power spectra is linear and truncated for emphasis.}
   \label{fig:filt_pow_kern_2}
\end{figure}

\citet{birch2004} showed how phase-speed filtering  affects the properties of   p-mode sensitivity
kernels for sound speed perturbations. Here we study how  f-mode kernels change as the spatial frequency content of the input f-mode power spectrum is varied.   To isolate particular f modes, we apply a multiplicative filter $G(k)$ to the wavefield,  in addition to the filter $F$ that contains the instrumental MTF and a mode-mass correction (\S~\ref{sec:uniform}). This additional multiplicative filter is a simple Gaussian function,
\begin{equation}
\label{gfilt}
G(k)=\exp{\left[\frac{-(k-k_0)^2}{2\sigma^2}\right]},
\end{equation}
with central wavenumber $k_0$ and standard deviation $\sigma$. Note that f modes with different wavenumbers probe different depths \citep[e.g.][]{duvall2000}.

Figure~\ref{fig:filt_pow_kern_2} shows examples of sensitivity kernels, using the filter $G(k)$ for  $k_0 R=487$, $900$, and $1322$ and  $\sigma R=348$ ($R=696$~Mm is the radius of the Sun). Plotted alongside each kernel is the corresponding filtered  power spectrum.  The particular part of the f-mode ridge that is isolated by the additional filtering is evident from these power spectra. As the bandwidth is reduced by filtering, the kernels show more fine structures and are more spread out in space (see Figure~\ref{fig:kerns10} for a comparison with a kernel computed with the full f-mode power).  As $k_0$ is varied from smaller to larger values, more and  more fine  structure  becomes apparent in the kernels.  We conclude that the sensitivity kernels must be tuned to account for the correct power spectrum before any high spatial resolution inversion is attempted, especially at scales comparable to the wavelength.


\subsection{Far-field approximation to the kernels}
\label{sec:farfield}

\begin{figure}[t]
   \includegraphics[width=\linewidth]{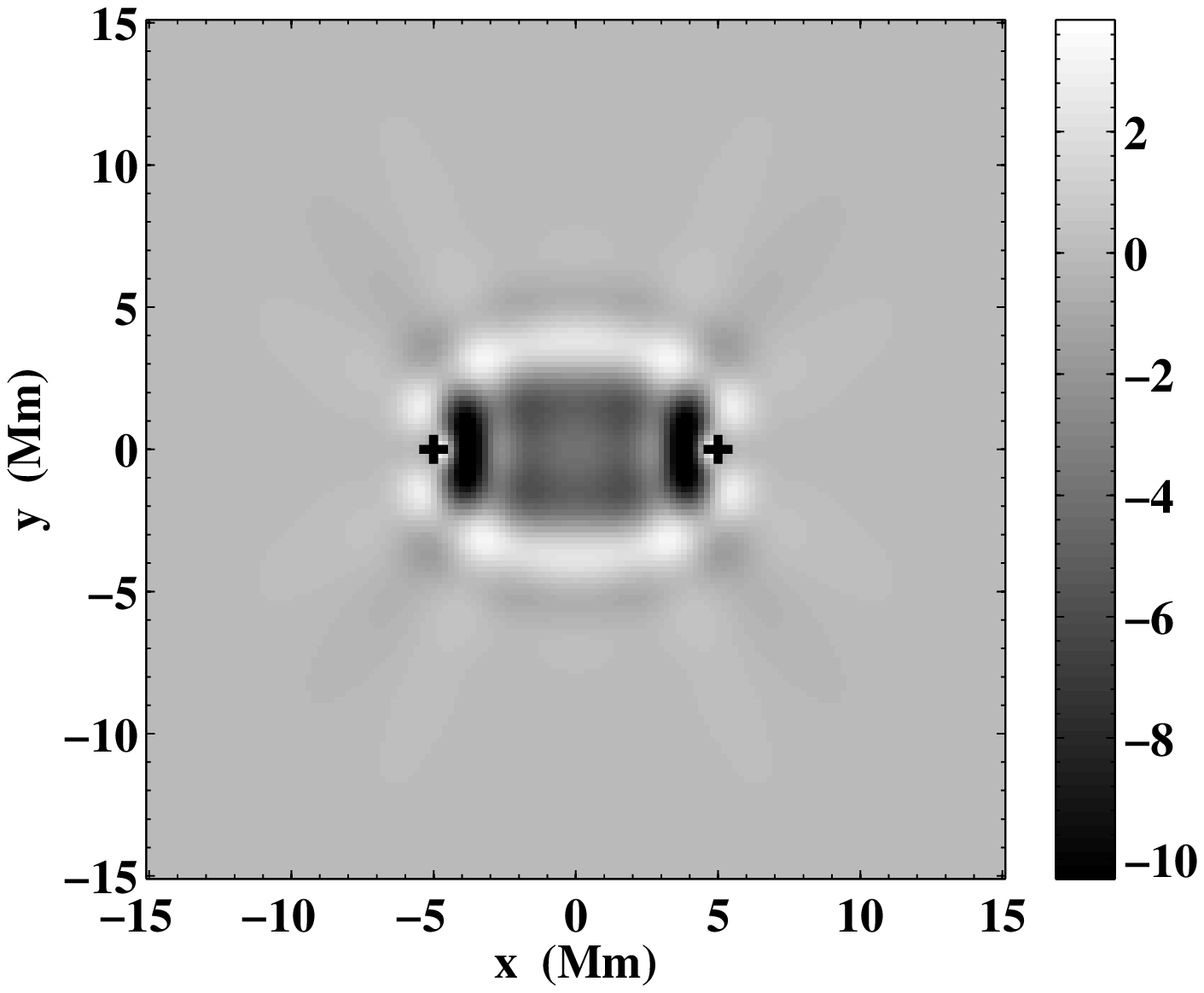}
\centering
   \caption{Far-field kernel ${\rm K}_x^{\rm diff}$ computed from equation~(\ref{farfieldkernel}) and discussed in \S~\ref{sec:farfield}. Note that this kernel is not defined at the two observation points,  given by the crosses. The spatial integral of this kernel is within $5$~\% of the spatial integral of the exact kernel (see Fig.~\ref{fig:kerns10}\textit{a}). The units are  s\,Mm$^{-2}$ (km/s)$^{-1}$. It is computed in just several seconds.}
   \label{fig:kern_farfield}
\end{figure}

\begin{figure}[t]
   \includegraphics[width=\linewidth]{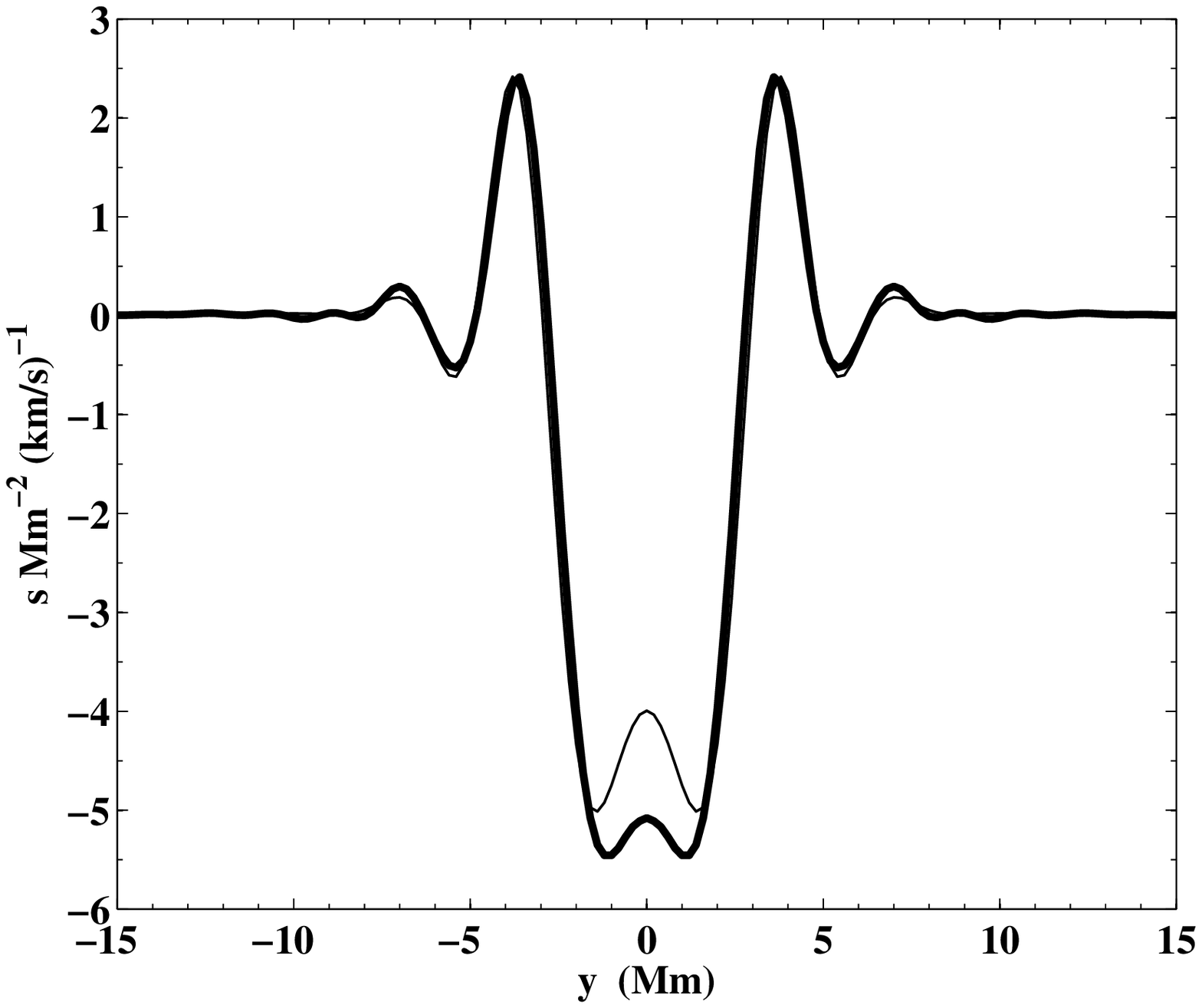}
\centering
   \caption{Cuts through the exact and far-field kernel along $x=0$. The
     heavy solid line is the cut through the kernel of
     Figure~\ref{fig:kerns10}\textit{a}, and the thin line is a cut through
     the far-field approximate kernel from Figure~\ref{fig:kern_farfield}.}
   \label{fig:cut_kern_farfield}
\end{figure}

We now consider an approximation to the sensitivity kernels given by equation~(\ref{K}). This is carried out in the far field, i.e., for points $\bx$ away from the observation points $\1$ and $\2$. The solution  will enable us to speed up the numerical computation of the kernels, as well as to expose some of the main features of the kernels that have already been discussed. The approximation allows for the analytical solution of the integrals given in equations~(\ref{int1})-(\ref{int4}). The details are presented in Appendix~\ref{app:far-field}, where the final expression for the approximate kernel is equation~(\ref{farfieldkernel}).

Figure~\ref{fig:kern_farfield} is an example  sensitivity kernel computed in the far-field approximation. It  resembles quite well the `exact' kernels of   \S~\ref{sec:ptp} (see Fig.~\ref{fig:kerns10}a).  The agreement between the exact and far-field kernel is everywhere within 10~\% except within a $1-2$~Mm radius about the two observation points. In fact, the far-field kernel is not defined at the observation points.  Figure~\ref{fig:cut_kern_farfield} shows one-dimensional cuts along the $x=0$ line of both the exact and approximate ${\rm K}_x^{\rm diff}$.

The far-field approximation kernel, equation~(\ref{farfieldkernel}), clearly brings out two important features. The first is the origin of the ellipses and hyperbolae mentioned in \S~\ref{sec:ptp}. Both the ellipses ($\Delta_1+\Delta_2=\Sigma={\rm const}$) and hyperbolae ($\Delta_1-\Delta_2=\Lambda={\rm const}$) are clearly present in the numerator of  expression~(\ref{farfieldkernel}). The second is the presence of damping  through the  $\kappa_\ii$ terms ($\kappa_{\rm i}={\rm Im}(\kappa)=\gamma\omega/g$). The damping causes the kernel to decay by a factor $\sim{\rm exp}[-\omega\gamma(\Sigma+\Delta)/g]$ as one moves away from the ray path, where $\gamma$ is the damping rate.

\subsection{Comments on the ray approximation}
\label{sec:ray}

Ray theory is a high-frequency approximation. According to \citet{kosovichev1997}, in the ray approximation the sensitivity of travel times to flows is confined to the ray path which joins the two observation points. In the case of f modes, the ray path is a straight line between the two observation points. 
For simplicity we study a configuration where the two observation points lie on the $x$ axis, $\1=(0,0)$ and $\2=(\Delta,0)$. The one-way travel-time perturbation becomes
\begin{equation}
\delta \tau^{\rm ray} = \frac{1}{\omega_0} \int_{0}^{\Delta} \delta k_x(x) \, \id x ,
\end{equation} 
where $\delta k_x$ is the local change in $x$ component of the wavevector  and $\omega_0=3$~mHz is the frequency for which the $k$-averaged power spectrum is maximum. For the case of f modes and a flow $u_x(x)$, the travel-time  difference $\tau_{\rm diff}$ can be approximated as 
\begin{equation}
\tau^{\rm ray}_{\rm diff} = -\frac{4 \omega_0^2}{g^2}\int_{0}^{\Delta} u_x \, \id x.
\label{ray}
\end{equation}

We would like to check whether the two-dimensional kernels derived in this paper reduce to the ray kernels when we compute the cross-path integral, i.e., when flows vary slowly in the $y$ coordinate.
In order to facilitate an analytic expression for the travel times, we begin from the far-field approximation of  ${\rm K}_x^{\rm diff}$,  equation~(\ref{farfieldkernel}). Following \citet{tong1998}, the integral of ${\rm K}_x^{\rm diff}$ over $y$ can be estimated by the method of stationary phase (sp). We obtain the following 
\begin{equation}
\tau^{\rm sp}_{\rm diff} = -\frac{4 \overline{\omega}^2}{g^2}\int_{0}^{\Delta} u_x \, \id x ,
\label{sp}
\end{equation} 
with the definition
\begin{equation}
\overline{\omega} (\Delta) \equiv \left( \frac{\int_0^{\infty}P_\psi^2\omega^2 e^{-2\kappa_{\rm i}\Delta}\id\omega}{2 \int_0^{\infty}P_\psi^2 e^{-2\kappa_{\rm i}\Delta}\cos^2(\kappa_{\rm r}\Delta-\pi/4) \id\omega} \right)^{1/2} ,
\end{equation}
where $P_\psi(\omega)$ is an average of the power spectrum over wavenumbers, and is explicitly defined in Appendix~\ref{app:far-field}. The functions $\kappa_{\rm r}(\omega)$ and $\kappa_{\rm i}(\omega)$ are the real and imaginary parts of the complex wavenumber at resonance.  For $\Delta>5$~Mm, we find that $\overline{\omega}\simeq 3.3$~mHz is a weak function of $\Delta$. At shorter distances however,   $\overline{\omega}(\Delta)$ varies fast with $\Delta$.  Only in the case when the power distribution $P_\psi(\omega)$  peaks around $\omega_0$,  at large distances $\Delta$, and in the absence of damping, do we have $\overline{\omega}=\omega_0$ and thus recover the ray approximation given by equation~(\ref{ray}).  In practice, damping cannot be ignored, and the naive ray kernel should be corrected by replacing $\omega_0$ by $\overline{\omega}(\Delta)$.  

\begin{figure}[t]
   \includegraphics[width=\linewidth]{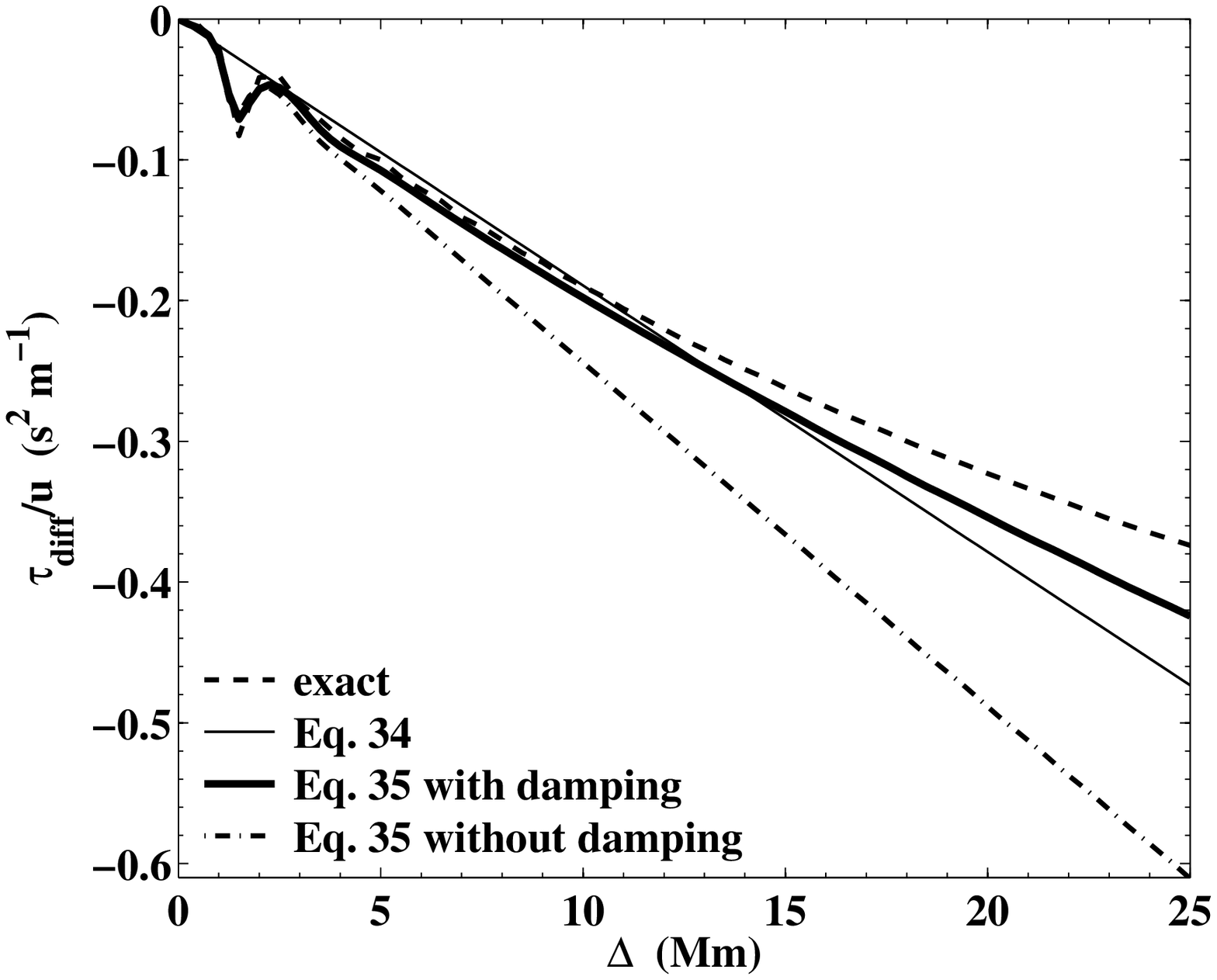}
\centering
   \caption{Scaled travel-time differences in the ray approximation, the stationary-phase approximation, and for the exact model. For each case, we consider a constant flow in the direction of $\1$ to $\2$. The exact solution (\textit{dashed line}) is for a flow  $u=200$~m\,s$^{-1}$ (see \S~\ref{sec:uniform}). The thin solid line corresponds to travel-time differences $\tau_{\rm diff}^{\rm ray}$ computed using equation~(\ref{ray}). The thick solid line is computed from equation~(\ref{sp}) and denotes $\tau_{\rm diff}^{\rm sp}$.  Finally, the dot-dashed line is the travel-time difference from equation~(\ref{sp}) without any damping.  The central angular frequency $\omega_0=3$~mHz.}
   \label{fig:tt_rayapprox}
\end{figure}

Figure~\ref{fig:tt_rayapprox} shows the travel-time differences for a uniform flow from equation~(\ref{sp}) for two cases: one with damping present  and one with damping set to zero. The difference between the two cases is quite large. Damping reduces the overall sensitivity to the flow (\textit{thick solid line}). The nonlinearity of the travel times with distance is a direct effect of the damping, and we see that turning off the damping restores linearity (\textit{dot-dashed line}).  For comparison we also plot the exact travel-time differences for a uniform flow of $200$~m\,s$^{-1}$, studied in \S~\ref{sec:uniform}. Remarkably, for $\Delta\lesssim 20$~Mm, the exact solution and the solution from equation~(\ref{sp}) are in good agreement, even though there have been two approximations made to arrive at equation~(\ref{sp}). The near-field wiggle at $\sim 2.5$~Mm is captured by the improved ray approximation.

\subsection{Sensitivity kernels for mean travel times}

\label{sec:mean}
\begin{figure*}[t]
   \includegraphics[ width=0.9\linewidth]{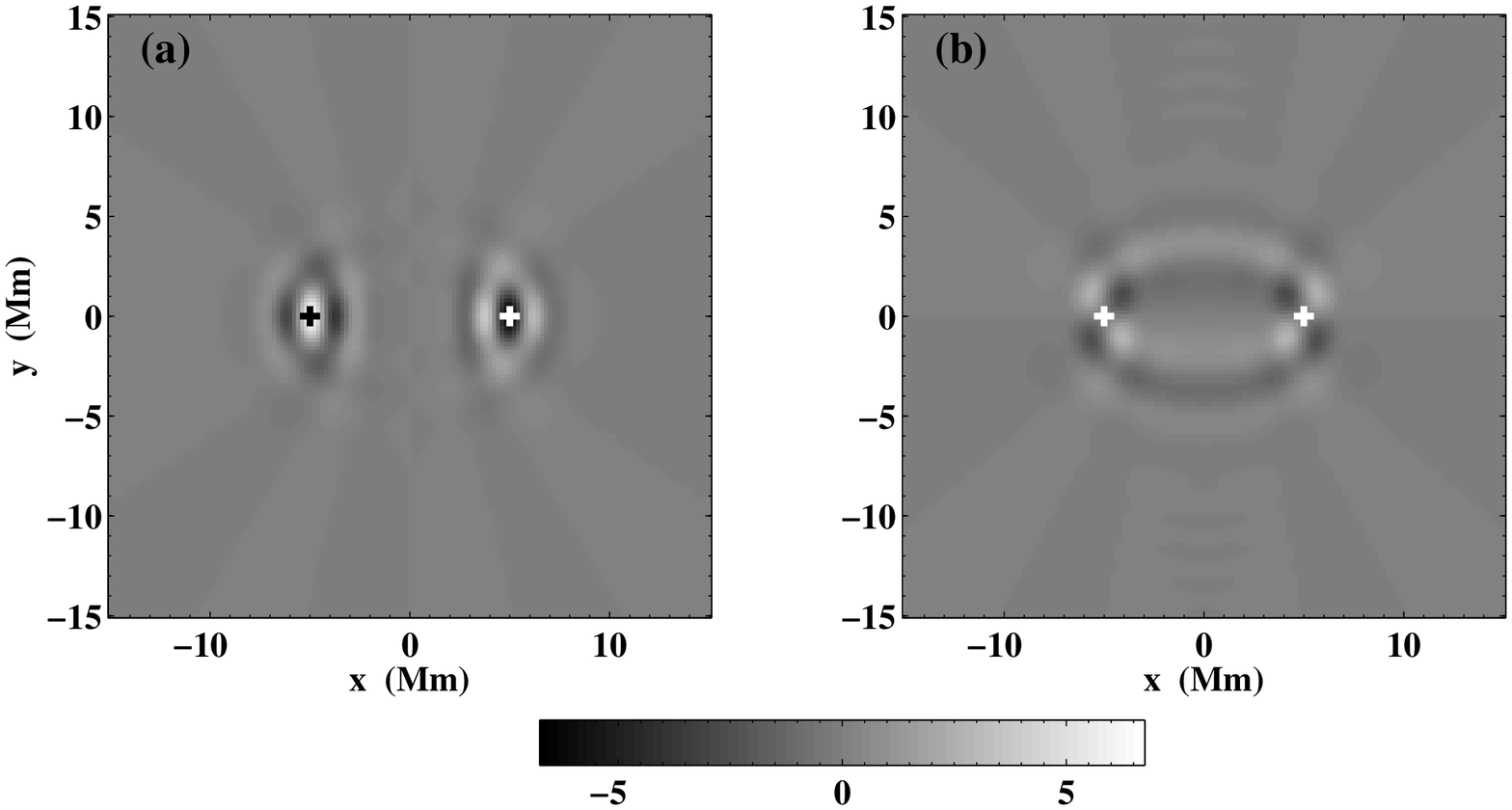}
\centering
    \caption{Sensitivity kernels for mean travel times $\tau_{\rm mean}$ discussed in \S~\ref{sec:mean},
      where (a) is ${\rm K}^{\rm mean}_x$ and (b) is ${\rm K}^{\rm mean}_y$ and $\Delta=10$~Mm.  The units are  s\,Mm$^{-2}$ (km/s)$^{-1}$.}
  \label{fig:kerns_mean}
\end{figure*}

Let us consider the mean travel times for point-to-point measurements  defined as
\begin{equation}
\tau_{\rm mean}(\1,\2)=\frac{1}{2}\left[\tau_+(\1,\2)+\tau_-(\2,\1)\right],
\end{equation}
where $\tau_\pm$ has been previously defined. 
It is sometimes assumed that mean travel times are not sensitive to flows. This is not strictly true, as we now discuss.

Following \citet{gizon2002} and \S~\ref{sec:ptp}, we may derive linear sensitivity kernels for mean travel times which satisfy
\begin{equation}
\label{tau-kmean}
\tau_{\rm mean}(\1,\2) = \doubint\bK^{\rm mean}(\1,\2;\bx)\cdot\bu(\bx)\,\id^2\bx,
\end{equation}
where 
\begin{equation}
\label{Kmean}
\bK^{\rm mean}=\frac{-2{\rm Re}\int_0^{\infty} {\rm Hilb}\left[\omega
  C^0(\bDelta,\omega)\right] \cC(\1,\2,\omega;\bx)\,\id\omega}{\int_0^{\infty}\omega'^2 |C^0(\bDelta,\omega')|^2 \, \id\omega'}.
\end{equation}
Here, Re takes the real part of the expression in the numerator and Hilb[$\dots$] denotes the Hilbert transform. 

To show how spatially-varying flows are related to  mean travel times $\tau_{\rm mean}$,  we compute
 the corresponding sensitivity kernels $\bK^{\rm mean}$ and plot them in
Figure~\ref{fig:kerns_mean}. First, we point out that the spatial integrals of ${\rm K}_x^{\rm mean}$ and ${\rm K}_y^{\rm mean}$ are both zero, indicating that a constant flow has no first-order effect on mean travel times.  
There are, however,  some structures  of significant magnitude in these kernels on small length scales (as large as for the travel-time difference kernel $\bK^{\rm diff}$).  Thus, flows that vary on a scale comparable to the wavelength may leave their signature in the mean travel times.

\section{Comparison with simplified three-dimensional kernels}
\label{sec:3D}

\citet{birch2007} computed three-dimensional kernels for the effect of local flows on travel-time differences, taking into account solar stratification but neglecting the advection of sources and damping.  As we have shown here, the effects of source and damping advection are, for practical purposes, negligible compared to the direct advection of the waves by the flow. Hence, it is meaningful to compare the two calculations.

The 3D kernels, denoted here as ${\bf K}^{\rm 3d}$, give the linear
sensitivity of travel-time differences to general three-dimensional
steady flows, ${\bf u}(\bx,z)$, and satisfy
\begin{equation}
\tau_{\rm diff}(\1,\2)=\int\!\!\!\!\int\!\!\!\!\int\! {\bf K}^{\rm
3d}(\1,\2;\bx,z)\cdot{\bf u}(\bx,z)\,\id^2\bx\,\id z .
\end{equation}
In order to compare the three-dimensional f-mode kernels, ${\bf K}^{\rm
3d}$, with the two-dimensional kernels computed in the previous sections, 
we define the depth-integrated kernel
\begin{equation}
{\bf K}^{\rm BG} = \int {\bf
K}^{\rm 3d}(\1,\2;\bx,z)\; \id z ,
\end{equation}
where the integral is over all $z$. The kernels  ${\bf K}^{\rm BG}$ give the sensitivity of travel-time differences to flows which are independent of depth.

\begin{figure*}[t]
  \includegraphics[ width=\linewidth]{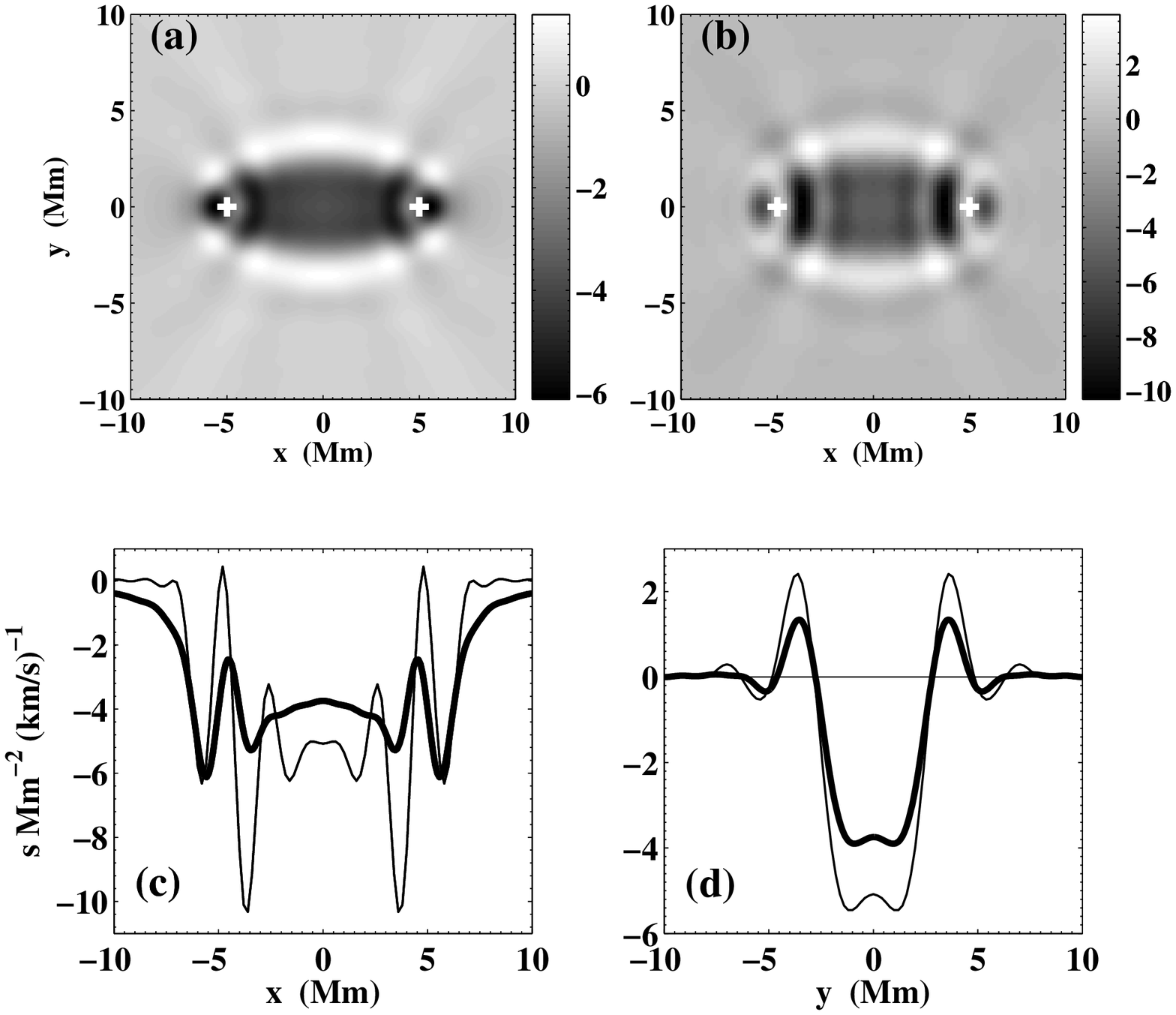}
\centering
   \caption{Comparison of the $x$ components of ${\bf K}^{\rm BG}$
\citep[the depth-integrated kernel from][]{birch2007} and the
corresponding two-dimensional f-mode kernel computed in this paper, $\bK^{\rm diff}$, for
the case $\Delta=10$~Mm.  Panel~(a) shows  ${\rm
K}_x^{\rm BG}$.  Panel~(b) shows  ${\rm K}_x^{\rm diff}$.  The horizontal integrals of the two
kernels agree to within a few percent. Panels (\textit{c}) and
(\textit{d}) show one-dimensional cuts through the ${\rm K}_x^{\rm BG}$
kernel (\textit{thick lines}) and the two-dimensional kernel ${\rm K}_x^{\rm diff}$
(\textit{thin lines}) along the $y=0$ and $x=0$ lines respectively.
In all panels the units of the kernels are s\,Mm$^{-2}$ (km/s)$^{-1}$.}
 \label{fig:3dkern_2dkern}
\end{figure*}

Figure~\ref{fig:3dkern_2dkern} compares an example depth-integrated
kernel, ${\bf K}^{\rm BG}$, with the corresponding two-dimensional
kernel from this paper.   The total integrals of the two kernels
agree to within a few percent.   Notice, however, that the details of
the fine structure are somewhat different.  These differences are likely
the result of the somewhat different assumptions regarding the source
covariance, the damping model, and the instrumental point-spread function.

The two bottom panels of  Figure~\ref{fig:3dkern_2dkern} show
one-dimensional cuts through the two kernels. The sizes of the first Fresnel
zones are equal; this shows that the dominant wavelengths in the two
calculations are similar. The kernel  ${\bf K}^{\rm BG}$ shows
somewhat less ringing than the kernel computed in this paper, which could be due to a slightly different frequency content.   The differences in the fine structure of these two
kernels demonstrate the importance of accurately modeling the true
bandwidth of solar f modes (see \S~\ref{sec:filtering}).

\section{Discussion}
\label{sec:discussion}

We have shown how to compute f-mode travel-time shifts due to uniform flows. The result depends on travel distance in a complicated way for several reasons: near-field effects at short travel distances and damping at large distances. In addition, we find that the linearization of the travel times with the flow amplitude is valid to within 10\% for flows with amplitudes less than about $250$~m\,s$^{-1}$ and travel distances less than about 25~Mm. For larger flows or larger distances, non-linear effects become important, as the travel times do not scale like the flow. This non linearity may complicate the interpretation of time-distance measurements because large-amplitude flows are observed on the Sun, including, e.g., rotation (hence the need to at least remove the main effect of rotation  before applying time-distance helioseismology). 

We computed the two-dimensional sensitivity of f-mode travel times to spatially-varying steady flows, using the first Born approximation (small flow regime). We found that the main physical mechanism responsible for travel-time shifts is, by far, the advection of the waves by the flow. The effects of the advection of wave sources (granulation) and wave damping by the flow are completely negligible. This will substantially simplify future computations of kernels, and justifies the omission of these terms in previous calculations. The kernels are also highly sensitive to the frequency content (central frequency, bandwidth) of the input f-mode power. Correctly tuning the model power spectrum to the observed one is therefore crucial for the use of these kernels in  any inversion. The point-to-point kernels computed here could easily be averaged to obtain point-to-quadrant or point-to-annulus kernels, a standard averaging technique used in time-distance helioseismology.

We showed that, at small spatial scales, the sensitivity kernels for Doppler velocity measurements depend very significantly on position on the solar disk and on the angle between the observation points and the center-to-limb direction. This seriously complicates the interpretation of travel-time measurements as the inversion problem is not a deconvolution. We note that close to the limb, kernels are nothing like ray theory would suggest. We caution that we have not considered the effects of foreshortening in this paper \citep[see][for a preliminary discussion]{jackiewicz2006}.

The two-dimensional f-mode kernels are fast to compute (especially in the far field) and compare well with  depth-integrated three-dimensional kernels. They may prove quite useful to infer near-surface horizontal flows.

\acknowledgements


\appendix 

\section{$\cC$ function}
\label{app:curlyc}
We would like to derive an expression for the function $\cC$ in equation~(\ref{curlyc}):
\begin{equation}
\label{app:deltac-curlyc}
\Delta
C(\1,\2,\omega)=\doubint \,\cC(\1,\2,\omega;\bx)\cdot\bu(\bx)\, \id^2\bx.
\end{equation}
First, we use the zero-order and scattered wave fields  in
 equations~(\ref{psi0}) and~(\ref{dpsi}) and plug them into
the first-order approximation to the cross-correlation, given by
\begin{equation}
\label{app:deltac}
\Delta
C(\1,\2,\omega)\approx E\left[\delta\psi^*(\1,\omega)\psi^0(\2,\omega)+
\psi^{0^{*}}(\1,\omega)\delta\psi(\2,\omega)\right],
\end{equation} 
where the flow $\bu$ is contained in the $\delta\psi$ terms. After some manipulation one can show upon comparison of
equation~(\ref{app:deltac-curlyc}) and~(\ref{app:deltac}) that

\begin{eqnarray}
\label{fullcurlyc}
\nonumber
\cC(\1,\2,\omega;\bx) &=&  \ii m_s \left[(\partial_\omega\kappa)K^*(\bDelta_2,\omega)\bnabla_{\bx}I(\bDelta_1,\omega)-(\partial_\omega\kappa^{*}) K(\bDelta_1,\omega)\bnabla_{\bx}I^*(\bDelta_2,\omega)\right]\\
&+& \ii (\partial_\omega m_s) \left[K^*(\bDelta_2,\omega)\bnabla_{\bx}K(\bDelta_1,\omega)-K(\bDelta_1,\omega)\bnabla_{\bx}K^*(\bDelta_2,\omega)\right],\\\nonumber
\end{eqnarray}
where we use the notation $\bDelta_1=\bx -\1$,
$\bDelta_2=\bx-\2$, and $\bDelta=\2-\1$ (see Fig.~\ref{fig:geometry} for the geometry schematic). The function $m_s$ is equivalent to the
source function in equation~(\ref{source}) for spatially uncorrelated sources, and
is related to the $m^0$ in equation~(51) of \citet{gizon2002}, where $m_s=\omega^2m^0/g^2$. In the above equation, we defined the following functions:
\begin{eqnarray}
I&=&\ell_z\cJ_0+\los_{\rm h}\cdot\cJJ_1,\\
K&=&\ell_z\tilde{\cJ}_0+\los_{\rm h}\cdot\tilde{\cJJ}_1,
\end{eqnarray}
where 
\begin{eqnarray}
\label{int1}
\cJ_0(\bx,\omega)&=&(2\pi)^5\kappa^* \doubint
\,e^{\ii\sbk\cdot\sbx}|G(k,\omega)|^2 F^*(\bk,\omega)\,\id^2\bk,\\
\tilde{\cJ}_0(\bx,\omega)&=&(2\pi)^2\kappa^*\doubint
\,e^{\ii\sbk\cdot\sbx}G^*(k,\omega)F^*(\bk,\omega)\,\id^2\bk,\\
\cJJ_1(\bx,\omega)&=&-(2\pi)^5 \ii\doubint
\, e^{\ii\sbk\cdot\sbx}\bk\, |G(k,\omega)|^2F^*(\bk,\omega)\,\id^2\bk,\\
\label{int4}
\tilde{\cJJ}_1(\bx,\omega)&=&-(2\pi)^2 \ii\doubint
e^{\ii\sbk\cdot\sbx}\bk\, G^*(k,\omega)F^*(\bk,\omega)\,\id^2\bk.
\end{eqnarray}
 The second line in
equation~(\ref{fullcurlyc}) is the Doppler shift, or advection, of the source
function due to the flow. This term is small compared to the wave advection term in the first line, as discussed in \S~\ref{sec:adv}.
We note that a complicated term has been left out of equation~(\ref{fullcurlyc}) that
contributes only at the observation points $\bx_1$ and $\bx_2$. We have calculated it, yet it is
negligeable for practical applications  and so we have ignored it. 


\section{Numerical computation of kernels}
\label{sec:computation}
We briefly point out some aspects of the numerical computation of the kernels. The most important step which determines the length of the computation is setting up the initial $x-y$ grid on which the kernels will be calculated.
The kernels of Figure~\ref{fig:kerns10}, which lie on a horizontal spatial grid of
resolution $0.2$~Mm in both $x$ and $y$ directions ($145\times 145$ points), are computed in $\approx
30$~minutes on a single processor machine, for a code written in
\verb!c++!. However, this value of the resolution turns out to be unnecessary. One
may use a much coarser grid, and then perform a two-dimensional Fourier interpolation to recover
the kernels onto a finer grid, if necessary. We have tested this and found that
the coarsest grid we may use for these kernels is roughly of a resolution of $\approx 1$~Mm. The
Fourier interpolation method is due to the Nyquist-Shannon sampling theorem. This theorem tells us that the minimum resolution necessary to recover the
full signal is
$\id x=\bar{\lambda}/4$, where $\bar{\lambda}$ is the dominant wavelength in the
problem, which, for f modes at power maximum, is about $5$~Mm. Implimentation of this speeds up
the computation quite effectively. The spacing for the $k$ and $\omega$ grids used for the computation are  approximately $\id\omega=7\times 10^{-5}$~rad s$^{-1}$ and $\id k=0.01$~Mm$^{-1}$, over the ranges $\omega=[1.5,5]$~mHz and $k=[0,2.5]$~Mm$^{-1}$.


\section{Far-field approximation to the kernels}
\label{app:far-field}

We would like to find an analytic approximation to the expression for the sensitivity kernels, starting from equation~(\ref{K}):
\begin{equation}
\label{K-W}
\bK^{\rm diff} =\frac{-2 {\rm Re}\int_0^{\infty}\ii\omega
    C^0(\bDelta,\omega) \,
    \cC(\1,\2,\omega;\bx)\id\omega}{\int_0^{\infty}\omega'^{2}|C^0(\bDelta,\omega')|^2 \id\omega'}.
\end{equation} 
We  consider for simplicity  sensitivity kernels where $\los=\hat{\bz}$, and we ignore the source advection
term in $\cC$ (line 2 of eq.~[\ref{fullcurlyc}]) , which we have already shown to be quite small.

Using the definition of the power spectrum (eq.~[\ref{pow}]), the zero-order cross-correlation (eq.~[\ref{cross-cov}]) can be written as
\begin{equation}
\label{app:c0}
C^0(\bDelta,\omega)=m_s(\omega)|\kappa|^2\doubint\frac{|F(\bk,\omega)|^2}{|k-\kappa|^2}e^{\ii\sbk\cdot\sbDelta}\,
\id^2\bk ,
\end{equation}
where $m_s(\omega)$ is the source function described in Appendix~\ref{app:curlyc}, $F$ is the filter function, and $\kappa$ is the complex wavenumber. Similarly, $\cC$ of equation~(\ref{fullcurlyc}) can be expressed as
\begin{eqnarray}
\label{app:curlycfull}
\cC =\frac{\ii m_s|\kappa|^2}{(2\pi)^2}\int\!\!\!\!\int\frac{\ii\bk'}{|k'-\kappa|^2} && \left[\frac{(\partial_{\omega}\kappa) F(\bk,\omega)F^*(\bk',\omega)}{k-\kappa}e^{-\ii\sbk\cdot\sbDelta_2+\ii\sbk'\cdot\sbDelta_1} \right. \nonumber \\
&& \left.
+\frac{(\partial_{\omega}\kappa^{*})F^*(\bk,\omega)F(\bk',\omega)}{k-\kappa^*}e^{\ii\sbk\cdot\sbDelta_1-\ii\sbk'\cdot\sbDelta_2}\right]\id^2\bk
\id^2\bk'.
\end{eqnarray}
The integrals over wavenumber in equation~(\ref{app:c0}) and (\ref{app:curlycfull})  can be performed analytically when the argument
of the exponential functions is large, i.e., in the far-field limit. For any smooth function $f(\bk)$ and horizontal vector $\bx$, we have, in the limit of large $x=\|\bx\|$ :
\begin{eqnarray}
\label{app:firstint}
\doubint_{-\infty}^{\infty}\frac{f(\bk)e^{\ii\sbk\cdot\sbx}}{k-\kappa}\id^2\bk&\approx&
\ii\frac{(2\pi)^{3/2}\kappa_{\rm r}}{\sqrt{\kappa_{\rm r}x}}f(\kappa_{\rm r}\hat{\bx})e^{\ii(\kappa_{\rm r}x-\pi/4)}e^{-\kappa_\ii x}\\\nonumber\\\nonumber\label{app:secondint}
\doubint_{-\infty}^{\infty}\frac{f(\bk)e^{\ii\sbk\cdot\sbx}}{|k-\kappa|^2}\id^2\bk&=&\frac{1}{2\ii\kappa_\ii}\doubint\frac{f(\bk)e^{\ii\sbk\cdot\sbx}}{k-\kappa}\id^2\bk-\frac{1}{2\ii\kappa_\ii}\doubint\frac{f(\bk)e^{\ii\sbk\cdot\sbx}}{k-\kappa^*}\id^2\bk\\\nonumber\\
&\approx&\frac{(2\pi)^{3/2}\kappa_{\rm r} e^{-\kappa_\ii x}}{2\kappa_\ii\sqrt{\kappa_{\rm r}x}}\left[f(\kappa_{\rm r}\hat{\bx})e^{\ii(\kappa_{\rm r}x-\pi/4)}+f(-\kappa_{\rm r}\hat{\bx})e^{-\ii(\kappa_{\rm r} x-\pi/4)}\right],
\end{eqnarray}
where $\kappa_{\rm r}={\rm Re}(\kappa)=\omega^2/g$ and $\kappa_\ii={\rm Im}(\kappa)=\gamma\omega/g$ (see \S~\ref{sec:surfacewaves}). 

Using these integral approximations we can now evaluate the 
kernel in equation~(\ref{K-W}). For simplicity, let the filter function be real and independent of $\bk$, and to simplify the notation let $F(k,\omega)\rightarrow F$. Integrating equation~(\ref{app:c0}) with the help of (\ref{app:secondint}) yields
\begin{equation}
\label{c0approx}
C^0(\bDelta,\omega)\sim \frac{P_{\psi} e^{-\kappa_\ii\Delta}}{\sqrt{\kappa_{\rm r}\Delta}}\cos(\kappa_{\rm r}\Delta-\pi/4),
\end{equation}
where $P_{\psi}(\omega)=(2\pi)^{3/2} m_s F^2(\kappa_{\rm r}, \omega) \kappa_{\rm r}^3/\kappa_\ii$ is approximately the power spectrum integrated over wavenumber. Note that this term is real. Similarly, from 
equation~(\ref{app:curlycfull}) one can show that
\begin{equation}
\label{curlycapprox}
{\rm Re}[\ii\omega\,\cC]=\frac{2\pi\, m_s\kappa_{\rm r}^5\,
  F^2 e^{-\kappa_\ii\Sigma}}{\kappa_\ii\sqrt{\Delta_1\Delta_2}}\left[\cos(\kappa_{\rm r}\Sigma-\pi/2)-\cos(\kappa_{\rm r}\Lambda)\right]\,\left(\hat{\bDelta}_1-\hat{\bDelta}_2\right),
\end{equation}
where $\Sigma=\Delta_1+\Delta_2$ and $\Lambda=\Delta_2-\Delta_1$, $\hat{\bDelta}_i$ denotes the unit vector in the direction of $\bDelta_i$, and we have used
$|\kappa|^2\approx\kappa_{\rm r}^2$
since the imaginary part is small (damping). Finally, plugging
equations~(\ref{c0approx}) and~(\ref{curlycapprox}) into equation~(\ref{K-W}) gives the far-field approximation to the sensitivity kernel:

\begin{equation}
\label{farfieldkernel}
\bK^{\rm diff}(\1,\2;\bx)\approx \frac{-2\left(\hat{\bDelta}_1-\hat{\bDelta}_2\right)}{\sqrt{2\pi
    g^5\Delta_1\Delta_2/\Delta}}\frac{\int_0^{\infty}P_{\psi}^2\omega^3
    e^{-\kappa_\ii\left(\Sigma+\Delta\right)}\left[\cos(\kappa_{\rm r}\Sigma-\pi/2)-\cos\kappa_{\rm r}\Lambda\right]\cos(\kappa_{\rm r}\Delta-\pi/4)\id\omega}{\int_0^{\infty}P_{\psi}^2
    e^{-2\kappa_\ii\Delta}\cos^2(\kappa_{\rm r}\Delta-\pi/4) \id\omega}.
\end{equation}
As noted in \S~\ref{sec:farfield}, the kernel is not defined at the observation points ($\Delta_1,\Delta_2=0$). A plot showing this kernel is given in Figure~\ref{fig:kern_farfield}. Expression~(\ref{farfieldkernel})  reduces to the result of ray theory in the infinite frequency limit, as demonstrated in \S~\ref{sec:ray}.


\end{document}